\documentclass[fleqn,usenatbib]{mnras}

\usepackage{newtxtext,newtxmath}

\usepackage[T1]{fontenc}
\usepackage{xcolor}

\DeclareRobustCommand{\VAN}[3]{#2}
\let\VANthebibliography\thebibliography
\def\thebibliography{\DeclareRobustCommand{\VAN}[3]{##3}\VANthebibliography}

\usepackage{graphicx}	
\usepackage{amsmath}	
\usepackage{etoolbox}

\title[Planetary Albedo of JWST Targets]{Constraining Planetary Albedo of JWST Targets in the TESS bandpass, using TESS, HST and \textit{Spitzer} Eclipse Depth Observations}

\author[Arora \& Goyal et al.]{
Rahul Arora,$^{1}$
and Jayesh Goyal$^{1}$\thanks{E-mail: jgoyal@niser.ac.in}
\\
$^{1}$School of Earth \& Planetary Sciences, National Institute of Science Education and Research (NISER), HBNI, Jatni, Khordha-752050, Odisha, India\\
}

\date{Accepted 2024 October 01. Received 2024 September 30; in original form 2024 February 16}

\pubyear{2024}

\begin{document}
\label{firstpage}
\pagerange{\pageref{firstpage}--\pageref{lastpage}}
\maketitle

\begin{abstract}
Albedo is one of the important characteristics of hot Jupiter exoplanets. However, albedo constraints have been obtained for very few exoplanets. In this work, we present the TESS Phase Curve observations of WASP-18b, WASP-19b, WASP-121b, WASP-43b, WASP-17b, and WASP-77b, all JWST  targets for atmospheric characterization and constrain their occultation depth as well as geometric albedo (A$_g$). We use a grid of self-consistent model atmospheres to constrain the metallicity, C/O ratio, and heat re-distribution for these six targets by fitting to their HST and/or \textit{Spitzer} observations and also compute the thermal contribution to total occultation depth in the TESS bandpass. We report the first value of TESS occultation depth for WASP-17b ($151_{-66}^{+83}$) and updated value for WASP-77Ab ($94_{-62}^{+53}$). We find self-consistent models constrain high values of thermal contribution to total occultation compared to Planck models. We find very low A$_g$ values for WASP-18b (< 0.089), WASP-19b (< 0.022), WASP-121b ($0.0^{+0.055}_{-0.104}$), WASP-77Ab ($0.017^{+0.126}_{-0.147}$) and significantly higher value for WASP-43b ($0.109^{+0.086}_{-0.088}$) and WASP-17b ($0.401^{+0.526}_{-0.307}$). We find WASP-17b lies in the ideal spot of low gravity and low equilibrium temperature, conducive for cloud formation, leading to high A$_g$. With the best-fit models, we constrain low heat re-distribution for all planets, with WASP-18b having the least. We also constrain sub-solar metallicity for all planets except WASP-17b and WASP-19b. We find a highly sub-solar C/O ratio for WASP-77Ab and WASP-43b, solar for WASP-18b, and super-solar for WASP-121b. The best-fit $P$-$T$ profiles show thermal inversion for WASP-18b and WASP-121b and none for WASP-77b and WASP-43b, which is in agreement with previous works.

\end{abstract}




\section{Introduction}
More than 5000 exoplanets have been detected, and this number is increasing daily. The atmospheric characterization of a few of these detected exoplanets has also started to unravel the physical and chemical processes in these faraway worlds. One such quantity that is important for the atmospheric characterization of exoplanets is their albedo, defined as the ratio of the light scattered by a planet to the light received by the planet. Since scattering of electromagnetic radiation is a wavelength-dependent process, albedo is also wavelength-dependent. Albedo can be an advantageous quantity to constrain the presence as well as the type of clouds, hazes, or any other sources of reflection in the atmosphere of planets or their surfaces. Moreover, constraining albedo can inform us about the presence of continents, oceans, and deserts on terrestrial planets. There are multiple definitions of albedo, but the most relevant one for this work is the geometric albedo. Geometric albedo $(A_g)$ is defined as the ratio of the planet's flux at zero phase angle (secondary eclipse) to the flux from a Lambert disk at the same distance and with the same cross-sectional area as the planet \citep{Seager2010}. 

Many measurements of the geometric albedo of individual exoplanets have been published till date, for example  HD 189733b and HD 209458b \citep{kren2023A&A...672A..24K}, KELT-20 b \citep{singh2024A&A...683A...1S}, HD 209458b \citep{brand2022A&A...659L...4B}, WASP-100b \citep{Jansen_2020}, HAT-P-32b \citep{Czesla2023A&A...677L..12C}, WASP-189b \citep{Deline2022} and WASP-43b \citep{Scandariato2022}. A few studies focused on a group of planets, for example \citet{Mallonn_2019} constrained albedos for hot and Ultra-hot Jupiter exoplanets, \citet{Wong_2020} and \citet{Wong_2021} focused on Year 1 and Year 2 TESS targets, respectively. \citet{pravH2023A&A...671L...3P} also provided evidence for temporal variability of albedo on KELT-1b . These measurements of albedo range from very low values, e.g. $A_g$ = 0.035 $\pm$ 0.045 for HD209458b \citep{Rowe_2008}, $A_g$ < 0.048 for WASP-18b \citep{Shporer_2019}, $A_g$ < 0.03 for WASP-104b \citep{Mo_nik_2018} indicating a more absorbing atmosphere, to higher values of $A_g$ = 0.40 (290-450 nm) for HD-189733 b  \citep{Evans2013}, $A_g$ < 0.24 for WASP-43b \citep{Keating_2017} and $A_g$ = 0.32$\pm$0.03 for Kepler-7b \citep{Demory2011}, indicating layers of clouds and presence of hazes/aerosols. \citet{Bla_ek_2022} presented optical geometric albedo constraints for five hot Jupiters, WASP-18b, WASP-36b, WASP-43b, WASP-50b and WASP-51b with Transiting Exoplanet Survey Satellite (TESS) observations. While they constrained the albedo for WASP-43b and WASP-18b using few self-consistent model spectra, the albedo from WASP-36b, WASP-50b, and WASP-51b were constrained assuming these planets are emitting like blackbodies. In this work, we use an extensive grid of self-consistent atmospheric models to constrain the albedos of six hot Jupiters, which have confirmed the allotted time for atmospheric characterization observations with the James Webb Space Telescope (JWST). The choice of these targets was also optimized based on the availability of TESS observations and their Hubble Space Telescope (HST) and \textit{Spitzer} observations. We in this work present the TESS bandpass secondary eclipse depth as well as the geometric albedo values of WASP-18 b \cite{2009Natur.460.1098H} , WASP-17 b \cite{Anderson_2009}, WASP-19 b \cite{Hebb_2009} , WASP-43 b \cite{wasp43}, WASP-121 b and WASP-77A b \cite{Maxted_2013}.

 Due to extremely close proximity to their host stars, most of the hot Jupiter exoplanets are expected to be tidally locked with their host star. Hence, by observing the transiting exoplanet at a phase angle of 0.5 (secondary eclipse), we can measure the flux received from the day-side hemisphere of the planet, providing us observations with sufficient Signal Noise Ratio (SNR) to constrain their albedo. Therefore, in this work, we focus on the geometric albedo of the day-side of the exoplanets. The flux that we measure from the day-side hemisphere comprises two components: the thermal emission from the planet due to its own temperature and the electromagnetic radiation reflected by the planet that it receives from its host star. The emission from FGK stars peaks in the optical wavelengths ($\sim$ 0.4 - 0.75 $\mu$m). Therefore, the reflected component will be maximum for hot Jupiters around FGK stars in this wavelength range. On the other hand, the emission from the planet can peak anywhere between 1 to 3 $\mu$m, depending on the atmospheric temperature structure. However, the emission from the planet and their reflected component are well separated in wavelength, thus allowing us to constrain the reflected component and, therefore, the albedo in optical wavelengths by constraining the planetary emission from near-infrared and infrared wavelength observations.  

 In this work, we utilize observations from TESS \citep{Ricker_2014} for measuring the secondary eclipse occultation depth from 0.6 to 1 $\mu$m (TESS bandpass). This TESS secondary occultation depth discussed in detail in Section \ref{sec:depth} comprises both the emission and reflected component. On the other hand, we use Hubble Space Telescope (HST) Wide Field Camera (WFC3) and \textit{Spitzer} observations to constrain the planetary emission. Thus, by extrapolating this measured planetary emission in the infrared wavelengths to optical wavelengths, using a grid of self-consistent simulated planetary atmosphere model spectra, we can subtract the planetary emission from the total day-side flux and thus constrain the reflected component and, therefore, the optical albedo of the planet. Section \ref{sec:depth} discusses this in more detail.

This paper is structured as follows. In Section \ref{sec:obs} we detail the facilities/instruments used for the observations and software/packages used to process the data. In Section \ref{sec:depth} we describe the details of the methodology and the governing equations used to compute geometric albedo using observed occultation depths and forward model grids. We benchmark our observed occultation depth with published values and present the comparison in Section \ref{sec:benchmark}. We detail our results for each planet in Section \ref{sec:reults}, discuss the results for the population of our targets in Section \ref{sec:disc}, and finally, we conclude in Section \ref{sec:conc}.

\section{Observation and data analysis}\label{sec:obs}
\subsection{TESS}
  TESS was launched in 2018 and operated in the optical and near-infrared wavelength region (0.6 - 1 $\mu$m). This work uses archived photometric data observed by TESS in multiple sectors. The archived data contains two sets of transit lightcurve data: Simple Aperture Photometry (SAP) and Pre-search Data Conditioning Simple Aperture Photometry (PDCSAP). The SAP data is corrected using the automatics TESS pipeline and corrected for the instrumental errors, which is further stored as the PDCSAP data. We use the PDCSAP dataset for this study as previously done in \cite{Bla_ek_2022} and \cite{Shporer_2019}. To further improve data quality, we performed a rudimentary cleaning of PDCSAP data by removing three sigma outliers from the TESS data and fitted the remaining systematics with ALLESFITTER as explained in Section \ref{subsec:allesfitter}. The Python-based \href{https://docs.lightkurve.org/reference/lightcurve.html}{LightKurve} package is used to process the PDCSAP data \citep{2016SPIE.9913E..3EJ}. All the archive data is available on  \href{https://mast.stsci.edu/portal/Mashup/Clients/Mast/Portal.html}{TESS MAST Archive}. For this study, we chose the data sets with 120 s of single exposure from each sector for the processing. The first step was to convert the time values from TESS default BJD(TBJD) to BJD$_{\rm{TBD}}$. Further data from all the sectors for each target planet were aligned sequentially and were fitted for the orbital period and epoch time. The fitting parameters were further used to fold the data from all available transits into a single normalized phase curve. The Light curve was cleaned for all the nan values to avoid errors. The resultant light curve was used for further analysis.

\subsection{Allesfitter}\label{subsec:allesfitter}
For the analysis of the TESS photometric data and constraining the planetary parameters, we use a photometric and radial velocity modeling and retrieval software package ALLESFITTER \citep{G_nther_2021}. ALLESFITTER performs retrieval with the help of two Bayesian sampling techniques, namely Markov chain Monte Carlo (MCMC) using emcee package \citep{Foreman_Mackey_2013} and nested sampling using dynesty package \citep{Speagle_2020}. The software also uses multiple other Python packages for photometric modeling like \href{https://github.com/pmaxted/ellc}{ellc}. The models generated by ALLESFITTER are further used to run a Bayesian analysis on the user's required free parameters. The package includes accounting for systematic noises and  Gaussian regression, which models for red noise or time-dependent noise such as stellar variability. In our study, the following free parameters were used: limb darkening coefficients (q$_1$ and q$_2$), the ratio of planetary radius to stellar radius $\Big(\frac{R_{\rm p}}{R_{\rm s}}\Big)$, sum of planetary and stellar radii divided by the semi-major axis $\Big(\frac{R_{\rm p} +R_{\rm s}}{a}\Big)$, cosine of inclination angle [cos(i)], epoch time/mid-transit time, orbital period of the planet, systematic noise of the instrument, A$_{\rm 1:beaming}$ (doppler boosting modulation), A$_{\rm 1:atmospheric}$ (Atmospheric boosting modulation), A$_{\rm 1:ellipsoidal}$ (ellipsoidal boosting modulation) and the brightness ratio J. We used nested sampling technique with 500 live points and a log(Z) threshold of 0.1, where Z is the Bayesian evidence. Each Bayesian modeling technique requires prior distribution for all the free parameters. In our light curve model we used Gaussian priors as our prior distribution function with the range of the distributions for each target obtained from Exomast\footnote{\url{https://exo.mast.stsci.edu}} for parameters like $\Big(\frac{R_{\rm p}}{R_{\rm s}}\Big)$,  $\Big(\frac{R_{\rm p} +R_{\rm s}}{a}\Big)$ and orbital period. Additionally,  prior range for other parameters were determined based on physical constraints such as 0 to 1 for Limb Darkening coefficients, cosine of inclination angle, A$_{\rm 1:beaming}$, A$_{\rm 1:atmospheric}$, A$_{\rm 1:ellipsoidal}$ and brightness ratio. Using the fitted parameters, ALLESFITTER calculates derived parameters like the radius of planet (R$_{\rm p}$), semi-major axis (a), inclination angle (i), impact parameter (b), eccentricity of the orbit (e), transit duration, epoch time of occultation, transit depth ($\delta_{\rm tra}$), occultation depth ($\delta_{\rm occ}$) and night side flux. For limb darkening, the surface brightness distribution I$_\lambda$(s, t) at any surface point s at time t incorporates established limb darkening laws U($\mu$), gravity darkening G(s, t), and the irradiation of the body by its companion (if any) H(s, t)U$_H$($\mu$)
given by

\begin{equation}
    I_\lambda(s, t) = I_0U(m)G(s, t) + H(s, t)U_H(m),
    \label{eq:limbfull}
\end{equation}

In this work we fit the limb darkening by a quadratic parameter fit given by

\begin{equation}
    U(\mu) = 1 - c_1(1-\mu) - c_2(1-\mu^2),
    \label{eq:limb}
\end{equation}

where, $c_1 = 2\sqrt{q_1}q_2$ and $c_2 = \sqrt{q_1}(1-2q_2)$.

\subsection{HST}
HST WFC3 G141 Grism secondary eclipse observations were used to constrain the infrared eclipse depth by fitting to grid of atmosphere model spectra. Among the six targets, the HST secondary eclipse observations were only available for four targets, WASP-18b \citep{Arcangeli_2018}, WASP-121b \citep{Evans_2017, Evans_2019}, WASP-77b \citep{Mansfield_2022} and WASP-43b \citep{Kreidberg_2014}. Therefore, only for these planets thermal emission contribution was constrained using the HST data.

\subsection{\textit{Spitzer}}
The \textit{Spitzer} satellite launched in 2003 and decommissioned in 2020 operated in the infrared wavelength regime. It started with 4 imagers IRAC 1, 2, 3, 4 operating at wavelengths $3.6 \mu $m, $4.5\mu$m,  $5.8\mu$m and $8\mu$m, respectively. Out of this four IRAC bands only two were functioning after 2009. HST and \textit{Spitzer} combined provide extended wavelength coverage in the infrared which can be very useful to constrain thermal emission component in the total occulation depth. \textit{Spitzer} $3.6 \mu $m, $4.5\mu$m eclipse depth observations have been used for five of our targets, WASP-18b \citep{Nymeyer_2011, Maxted_2013}, WASP-121 \citep{Evans_2017}, WASP-77b \citep{Mansfield_2022}, WASP-43b \citep{Blecic_2014} and  WASP-19b \citep{Garhart_2020}. For WASP-17b, $3.6 \mu$m and $8\mu$m eclipse depths were obtained from \citet{Anderson_2011}.

\section{Methodology}

\subsection{Occultation depth}
\label{sec:depth}

As mentioned in the introduction, we measure total occultation depth ($\delta_{\rm occ,obs}$) of a transiting exoplanet in the TESS bandpass. This total measured occultation depth  comprises of two components, the thermal occultation depth ($\delta_{\rm occ,th}$) which is observed due to masking of the thermal emission of the planet by the star and the reflected occultation depth ($\delta_{\rm occ,re}$) due to stellar flux reflected by the planet, i.e, 

\begin{equation}
    \delta_{\rm occ,obs} = \delta_{\rm occ,th} + \delta_{\rm occ,re} .
    \label{eq:1}
\end{equation}

Our this work aims to constrain the geometric albedo of the planet, by constraining the reflected component. The geometric albedo is defined as the amount of radiation relative to that from a flat Lambertian surface which is an ideal reflector at all wavelengths. To obtain the albedo the total occultation depth needs to be separated into thermal and reflected occultation depth. Theoretically the thermal contribution to occultation depth ($\delta_{ \rm occ,th}$) in the instrument bandpass can be calculated using:
\begin{equation}
    \delta_{\rm occ,th} =\Bigg(\frac{R_{\rm p}}{R_{\rm s}}\Bigg)^2\frac{B_{\rm int}(T_{\rm eq,p})}{B_{\rm int}(T_{\rm star})},
    \label{eq:2}
\end{equation}
 where R$_{\rm p}$ is radius of planet, R$_{\rm s}$ is the radius of star and 

 \begin{equation}
    B_{\rm int}(T_{\rm eq,p}) = \frac{\int_{\lambda_{\rm min}}^{\lambda_{\rm max}} B (T_{ \rm eq,p}, \lambda) \lambda R(\lambda)d\lambda}{\int_{\lambda_{\rm min}}^{\lambda_{\rm max}} \lambda R(\lambda)d\lambda} 
    \label{eq:3}
\end{equation}
 
 is the integrated blackbody flux at planet's equilibrium temperature ($T_{\rm eq,p}$) obtained by integrating the wavelength ($\lambda$) dependant blackbody flux $B (T_{\rm eq,p}, \lambda)$ over the instrument band-pass. $\lambda_{\rm min}$ and $\lambda_{\rm max}$ are the minimum and maximum wavelength of the bandpass, respectively,  and $R(\lambda)$ is the instrument response function. 

 $B_{\rm int}(T_{\rm star})$ is the integrated blackbody flux at the stellar effective temperature, calculated in the same way as that for the planet in equation \ref{eq:3}. 

 Assuming exoplanets as blackbody radiators can be a crude assumption. Due to presence of many radiative absorbing and emitting species in the planet's atmosphere, it could have a non-isothermal Pressure-Temperature ($P$-$T$) structure, thus making its spectral thermal emission different from that of a blackbody \citep{Evans_2017}. Therefore, in this work along with the computation of model thermal occultation depth assuming planet as a blackbody, we also compute the thermal occultation depth using the planet's emission spectrum computed using 1D radiative-convective equilibrium (RCE) planetary atmosphere model consistent with equilibrium chemistry. Using this RCE model emission spectrum will allow us to accurately determine the thermal contribution to the total occultation depth, thereby accurately constraining the reflected contribution and therefore the planetary albedo. RCE model emission spectrum used to remove the thermal contribution is the best fit model to the HST and \textit{Spitzer} observations of the planet, explained in detail in the next section. 
 
 The reflected occultation depth ($\delta_{\rm occ,re}$) can then be used to constrain the geometric albedo ($A_g$) using the following equation, 
 \begin{equation}
 \delta_{\rm occ,re} = A_g \Bigg(\frac{R_p}{a}\bigg)^2,
 \label{eq:4}
 \end{equation}
where a is the semi major axis of the planet \citep{winn2010transits}.

\subsection{Model Thermal Occultation Depth}
\label{sec:depth2}
To obtain a more realistic (non-blackbody) thermal occulation depth in the TESS bandpass we use a grid of exoplanet emission spectra computed using radiative-convective equilibrium $P$-$T$ profiles consistent with equilibrium chemistry \citep{Goyal2020, Goyal2021}. As compared to grid presented in \citet{Goyal2021} we now use a more dense grid for this work. Dense grid implies we have now reduced the spacing between subsequent grid parameters, with the aim to obtain more precise constraints if permitted by the observations. Our parameters for the grid are the same with addition of the internal temperature (T$_{\rm int}$) of the planet as one more grid parameter. This new grid has also been computed using  ATMO a 1D-2D radiative-convective equilibrium model for planetary atmospheres \citep{Tremblin2015, Drummond2016, Goyal2018}.However, we note that in this work we only use ATMO 1D simulations. To compute radiative-convective equilibrium (RCE) $P$-$T$ profiles we initialise the model with an guess profile. Equilibrium chemical abundances are then computed using Gibbs energy minimization based on this initial guess profile. Following this the total opacity is computed in each layer of the atmosphere, using the equilibrium chemical abundances and the respective Correlated-K opacities. This process is iteratively done under the constraints of hydrostatic equilibrium and conservation of energy, until we obtain a $P$-$T$ profile in radiative-convective equilibrium consistent with equilibrium chemical abundances. The radiative transfer equation is solved using the Discrete-Ordinate method and the convection is parameterized using the mixing length theory. More details and governing equations could be found in \citet{Drummond2016} and \citet{Goyal2020}.  

The grid of atmospheric profiles and emission spectra has been computed for 8 re-circulation factors (0.2 - 1.33), 22 metallicities (0.01x - 100x), 14 C/O ratios (0.01 - 2) and 4 T$_{\rm int}$ values (100\,K - 400\,K). Re-circulation factor (RCF) of "1.33" implies day-side profile without any re-circulation, corresponding to the theoretically maximum possible flux on the day-side of the planet's atmosphere. We note that this has been updated from maximum re-circulation factor of 1  in our previous grid, following \citet{Burrows2008}. \citet{Burrows2008} pointed out that the spectrum of the planet observed close to the secondary eclipse should be biased towards higher flux than we obtain by setting RCF as 1 (equivalent of f=1/2 in \citet{Burrows2008} with incidence angle of 60 degrees). This flux should be 2/3 times higher than what we obtain from RCF$_{\rm old}$ = 1 and incidence angle of 60 degrees, thus giving us the maximum value of RCF$_{\rm new}$ as 1.33 (RCF$_{\rm new}$ = RCF$_{\rm old}$/cos($\theta$)*2/3). 

The new model grid for the target planets have been computed at the spectral resolution of R$\sim$1000 with correlated-k opacities. We included H$_{2}$O, CO$_2$, CO, CH$_4$, NH$_3$, Na, K, Li, Rb, Cs, TiO, VO, FeH, CrH, PH$_3$, HCN, C$_{2}$H$_{2}$, H$_{2}$S, SO$_{2}$, H$^-$, and Fe opacities in our model. The details of the opacities, their treatment and model choices for computing the grid are all detailed in \citet{Goyal2020}. The model stellar flux for the stellar host of the target planet is obtained from the BT-Settl\footnote{\url{https://phoenix.ens-lyon.fr/Grids/BT-Settl/AGSS2009/SPECTRA/}} stellar model database \citep{Rajpurohit2013}. 

We fit this model grid to the HST and/or \textit{Spitzer} emission spectra observations of the target planets as described in \citet{Goyal2021}. The observed spectra were obtained from published studies referenced in Section \ref{sec:reults}. The best fit model with the lowest $\chi^2$ for each planet is used to obtain the thermal occultation depth in the TESS bandpass by integrating model planet and stellar flux convolved with the instrument response function over the TESS bandpass as shown in equation \ref{eq:3} (replacing blackbody flux with model flux). Subtracting this thermal occultation depth from the total TESS measured occultation depth, we obtain the reflected component and therefore constrain the geometric albedo in the TESS bandpass. We also compute 1$\sigma$ range of the thermal contribution using the model spectra that lies within 1$\sigma$ $\chi^2$ of the best fit model $\chi^2$ when fitted to the observed spectra. The errors on both, the measured total occultation depth and thermal contribution from the model are propagated to calculate the errors on albedo, all shown in Table \ref{tab:model_fit_param}. By fitting this self-consistent model grid to HST and/or \textit{Spitzer} observations of target planets, we are also able to obtain constraints on the energy re-circulation, metallicity and C/O ratio of these planets, as shown in Table \ref{tab:model_fit_param}.

\section{Benchmarking}\label{sec:benchmark}
We benchmark our analysis of TESS observations with already published results. We use WASP-18b as our benchmarking target. Full phase curve and secondary eclipse of WASP-18b was analysed in  \cite{Shporer_2019} and \cite{G_nther_2021}. WASP-18 b was observed by TESS in 4 different sectors 2,3,29 and 30 in the year 2018, 2018, 2020 and 2020, respectively. Throughout the observations, TESS observed 26, 20, 21 and 24 transits in the 4 sectors, respectively. In our current work we use data from all four sectors and phase fold them to obtain a phase curve with higher SNR. We then used ALLESFITTER to fit for the occultation depth as described in Section \ref{sec:obs}. We obtain an occultation depth of 337 $\pm$ 12 ppm, consistent with 342 $\pm$ 15 ppm as reported by \citet{G_nther_2021}, 345 $\pm$ 11 ppm by \citet{Bla_ek_2022} and 341 $\pm$ 18 ppm by \citet{Shporer_2019}. All the differences that we see are within the error bars of each measurement. We also constrain R$_{\rm p}$/R$_{\rm s}$ to 0.09768 $\pm$ 0.00019, consistent with 0.09757 $\pm$ 0.00014 from \citet{G_nther_2021} and in close agreement with 0.09716 $\pm$ 0.00014 from \citet{Shporer_2019}. We note that there can be differences in the transit depth obtained by squaring the fitted R$_{\rm p}$/R$_{\rm s}$ and the derived transit depth as highlighted by \citet{Heller2019}. The derived transit depth for WASP-18b that we constrain in this work is 10664 $\pm$ 16 ppm consistent with 10665 $\pm$ 18 ppm from\footnote{\url{https://github.com/MNGuenther/allesfitter/blob/master/paper/WASP-18/allesfit_sine_physical/results/mcmc_derived_corner.jpg}} \citet{G_nther_2021}. This derived transit depth value differs from that quoted in \citet{Shporer_2019} ($9439_{-26}^{+27}$ ppm) as they compute transit depth by squaring the R$_{\rm p}$/R$_{\rm s}$.

\section{Results}\label{sec:reults}
We showed benchmarking of our analysis of the TESS observations for WASP-18b in the previous section. We then analysed the TESS observations of other target planetary systems in our list. We constrain their R$_{\rm p}$/R$_{\rm s}$, occultation and transit depth, and many more fitted and derived parameters, all listed in Table \ref{tab:allesfitter_fitted} and Table \ref{tab:allesfitter_derived}. We also fit published HST and/or \textit{Spitzer} observations of the target planets to our self-consistent grid of model spectra described in Section \ref{sec:depth2} and thus constrain their RCF, metallicity, C/O ratio and internal temperature, all shown in \ref{tab:model_fit_param}. Using this best-fit model spectra for each planet, 
we compute the thermal contribution to the total measured occultation depth in the TESS bandpass and constrain the geometric albedo of the planets, all listed in Table \ref{tab:albedo}. For comparison, we also compute the thermal contribution using a black-body Planck function at the equilibrium temperature of the planet and thus constrain the geometric albedo, which is also shown in Table \ref{tab:albedo}. In this section, we detail the results that we obtain for each planet.

\subsection{WASP-18b}

\begin{figure}
    \centering
    \includegraphics[scale=0.28]{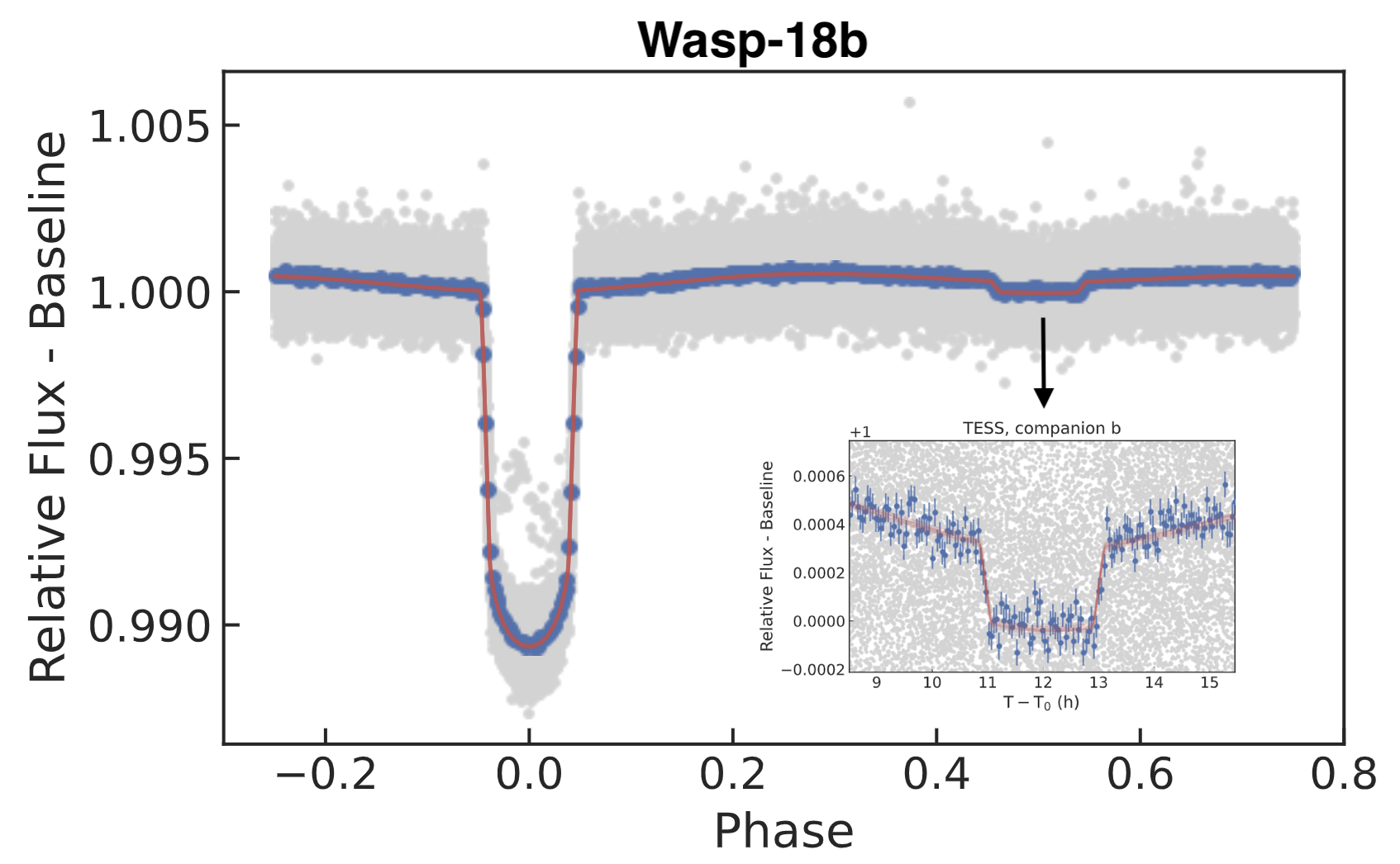}
    \caption{TESS full phase curve fit for WASP-18b.}
    \label{fig:pcwasp-18}
\end{figure}

\begin{figure}
    \centering
    \includegraphics[scale=0.5]{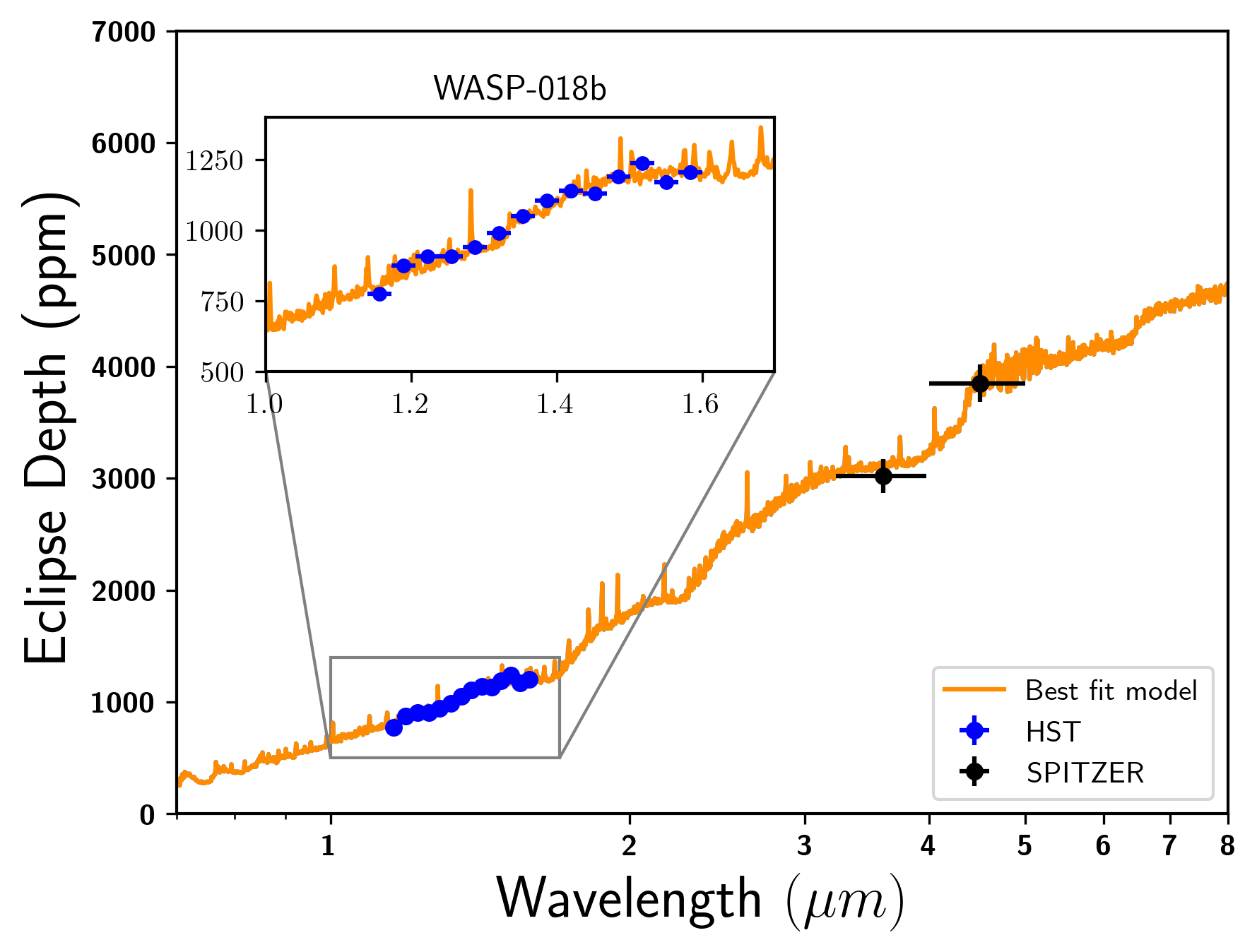}
    \caption{Best-fit emission spectrum model fitted to HST and \textit{Spitzer} observations of WASP-18b. The spectral region covered by HST is also zoomed in the Figure.}
    \label{fig:bestfit_wasp-18}
\end{figure}

\begin{figure*}
    \centering
    \includegraphics[scale=0.45]{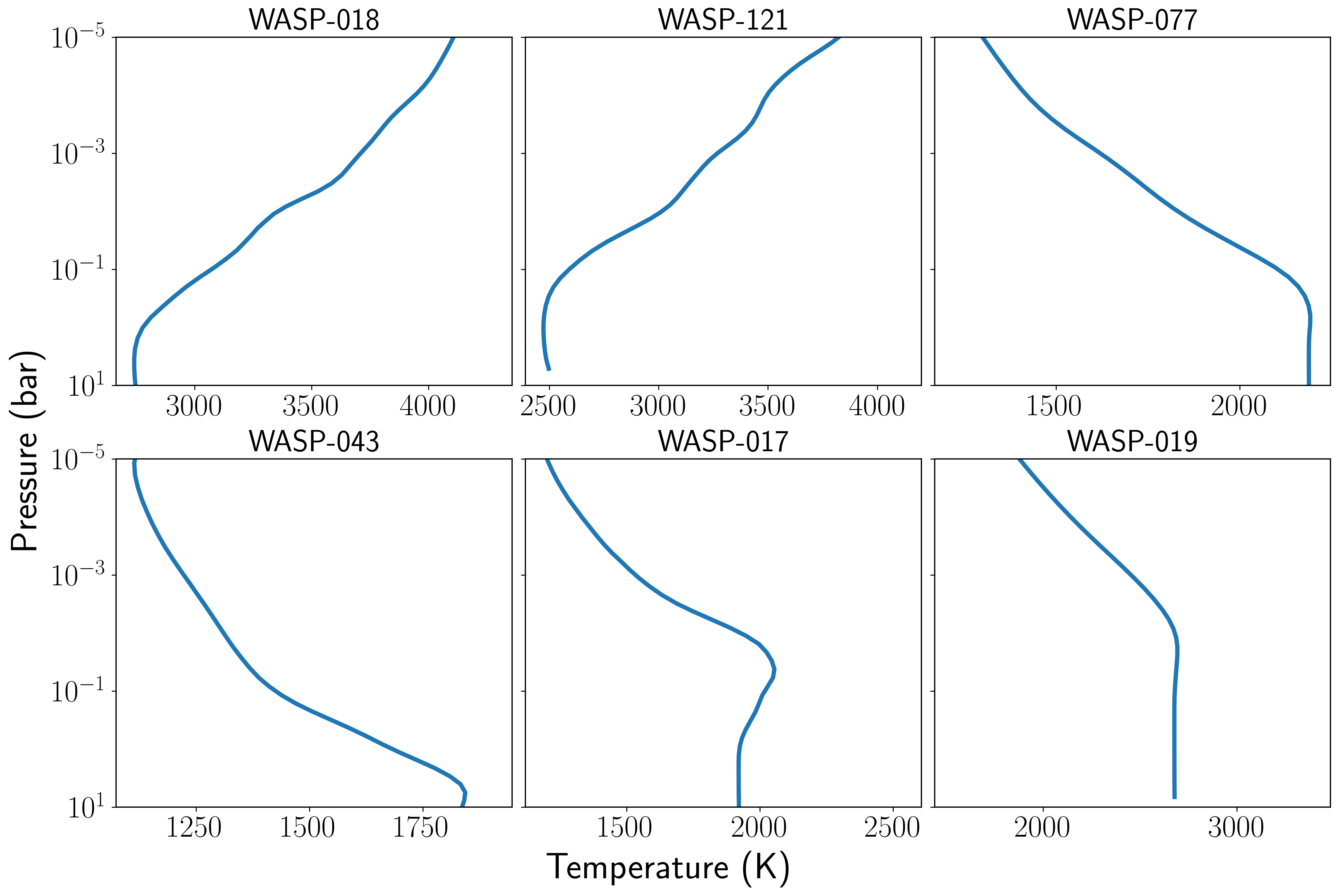}
    \caption{\textsf{Best-fit self-consistent radiative-convective equilibrium $P$-$T$ profile for each of our target planets.}}
    \label{fig:pt}
\end{figure*}

With the rich dataset of WASP-18b as described in the bench-marking section, we were able to constrain its TESS occultation depth to 337 $\pm$ 12 ppm with 28 $\sigma$ detection significance. We analysed the thermal contribution to the total TESS occultation depth of WASP-18b, with both theoretical Planck function and fitting a grid of self-consistent model atmospheres to previous HST \citep{Arcangeli_2018} and Spitzer \citep{Maxted_2013} observations . Using the Planck function at the equilibrium temperature to compute the thermal contribution, we constrain the geometric albedo to 0.297. The best-fit self-consistent model that we obtain for WASP-18b using its HST and \textit{Spitzer} observations as shown in Figure \ref{fig:bestfit_wasp-18} has a RCF value of 1.2, metallicity of 0.63x solar, solar C/O ratio (0.55), with the reduced $\chi^2$ value of 1.28, indicating good fit to the observations. High value of RCF corresponds to very poor re-circulation in the planet's atmosphere. The best-fit model also corresponds to a $P$-$T$ profile with a thermal inversion shown in Figure \ref{fig:pt}, thus confirming the presence of thermal inversion in the atmosphere of WASP-18b as found by \citet{Arcangeli_2018, Coulombe2023}. Using this best fit model spectra we obtain a thermal contribution of $387^{+107}_{-6}$ ppm in the TESS bandpass, leading to albedo value of <0.089. The predicted thermal contribution of \citet{Bla_ek_2022} was also higher than the measured occultation depth, using a completely different atmospheric model. Even with 3 $\sigma$ upper limit on total occultation depth (373ppm) and the thermal contribution of 375 ppm that we obtain as the lowest thermal contribution among the models within the 3 $\sigma$ of the best-fit model, the albedo is constrained to -0.0025. In summary, the TESS measured occultation depth for WASP-18b could be explained by thermal emission alone, without any reflected component with self-consistent models. However, using the Planck function which is a very crude approximation for these planets, leads to substantial reflected component. With the TESS phase curves we also constrain nightside flux for WASP-18b to be 38 $\pm$ 12 ppm, which is substantially lower than the day-side flux.

\subsection{WASP-121b}

\begin{figure}
    \centering
    \includegraphics[scale=0.5]{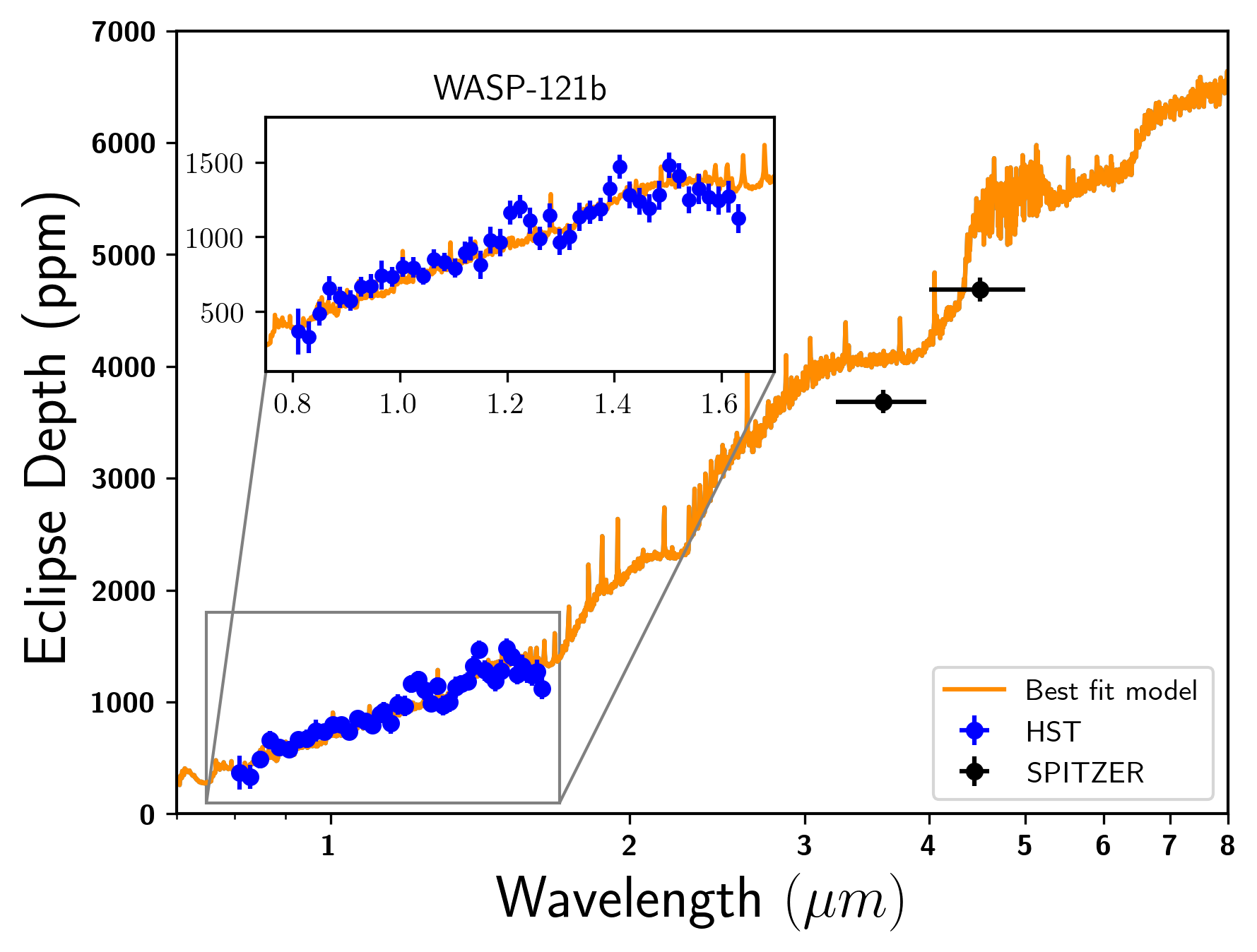}
    \caption{Best-fit emission spectrum model fitted to HST and \textit{Spitzer} observations of WASP-121b.}
    \label{fig:bestfit_wasp-121}
\end{figure}

For WASP-121 b TESS archive has 3 sectors of data (Sector 7, 33, 34) taken in the year 2019, 2020, 2021 respectively. We contrained the planet to stellar radius ratio to be  $0.12181_{-0.00030}^{+0.00032}$ which is lower than $0.12355_{-0.00033}^{+0.00029}$ published by \citet{Bourrier2020},  $0.1230_{-0.00025}^{+0.00025}$ by \citet{Yang_2022} and $0.12488_{-0.00030}^{+0.00029}$ by \citet{Daylan_2021}. The reason for this discrepancy could be attributed to the datasets used in these studies, we have used all the available sector data while \citet{Bourrier2020}, \citet{Yang_2022} and \citet{Daylan_2021} have used data from only one sector. \citet{Bourrier2020} and \citet{Daylan_2021} have used only sector 7 data. We constrain the occultation depth to 405 $\pm$ 22 ppm with 18.4 $\sigma$ detection significance as compared to $419_{-41}^{+47}$ ppm by \citet{Bourrier2020} and $482_{-41}^{+39}$ ppm by \citet{Daylan_2021}. The best-fit self-consistent model that we obtain for WASP-121b using its HST-G141 \citep{Evans_2017}, HST-G102 \citep{Evans_2019} and \textit{Spitzer} \citep{Garhart_2020} observations as shown in Figure \ref{fig:bestfit_wasp-121} has a RCF value of 1.0, metallicity of 0.63x solar, C/O ratio of 0.8, with the reduced $\chi^2$ value of 1.56, indicating good fit to the observations. The value of RCF is also quite high for WASP-121b indicating poor re-circulation in the planet's atmosphere but less than that of WASP-18b. For WASP-121b too the best-fit model corresponds to a $P$-$T$ profile with a thermal inversion shown in Figure \ref{fig:pt}, thus confirming the presence of thermal inversion in the atmosphere of WASP-121b as found by \citet{Evans_2017}. Using this best fit model spectra we obtain a thermal contribution of $405^{+36}_{-88}$ in the TESS bandpass, leading to albedo value of $0.0^{+0.055}_{-0.104}$, as compared to $0.07_{-0.040}^{+0.037}$ reported by \citet{Daylan_2021}. We also a constrain a 3 $\sigma$ upper limit of 0.063 on the geometric albedo with a total occultation depth of 471 ppm. As for WASP-18b, for WASP-121b also we obtain a substantial albedo of 0.268 when using a Planck function, while we obtain practically zero reflected component when we use self-consistent model grid. With the TESS phase curves we constrain nightside flux for WASP-121b to be $98_{-18}^{+22}$ ppm, which is substantially lower then the day-side flux similar to WASP-18b.

\subsection{WASP-77Ab}

\begin{figure}
    \centering
    \includegraphics[scale=0.28]{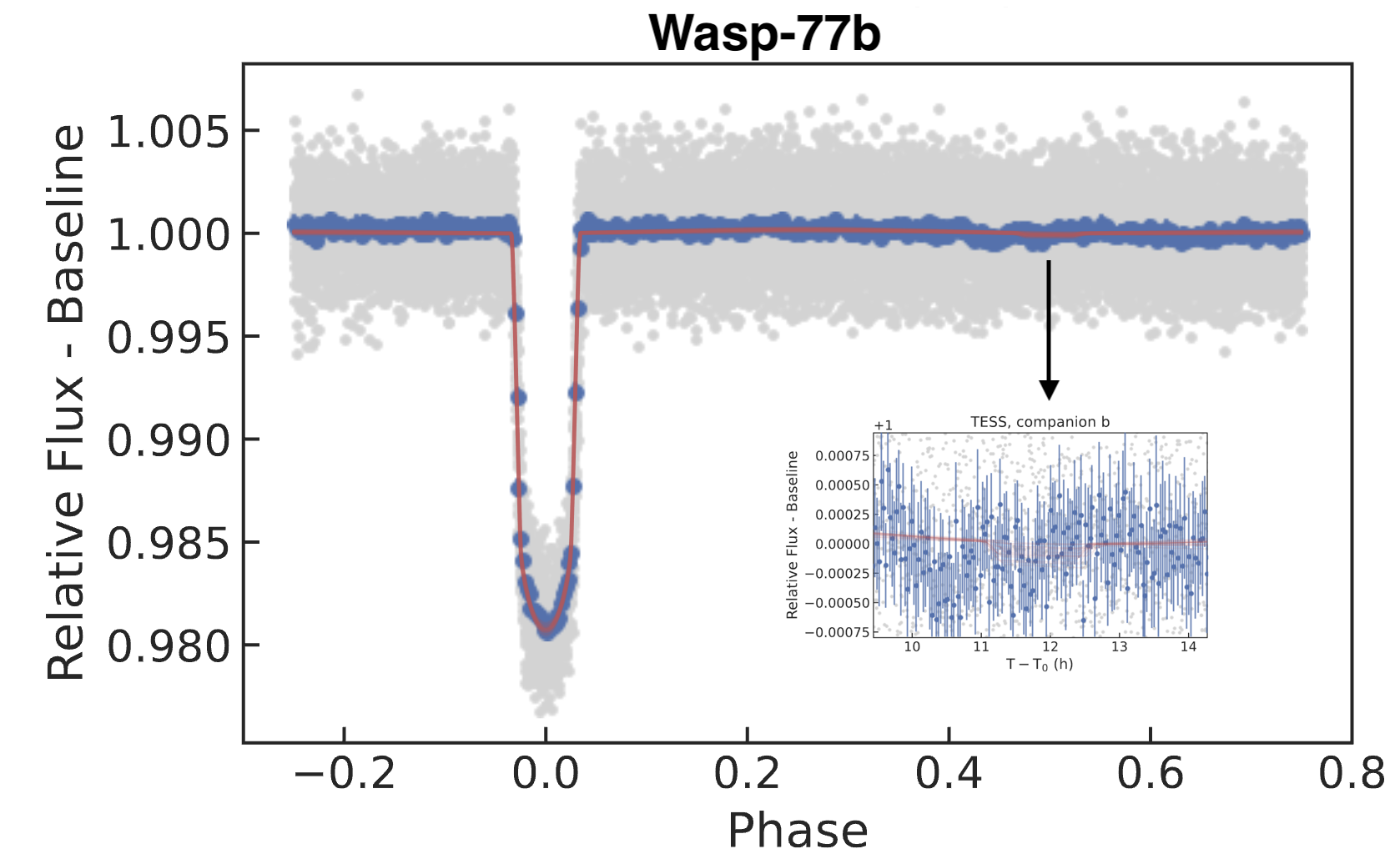}
    \caption{TESS full phase curve fit for WASP-77b.}
    \label{fig:pcwasp-77}
\end{figure}

\begin{figure}
    \centering
    \includegraphics[scale=0.5]{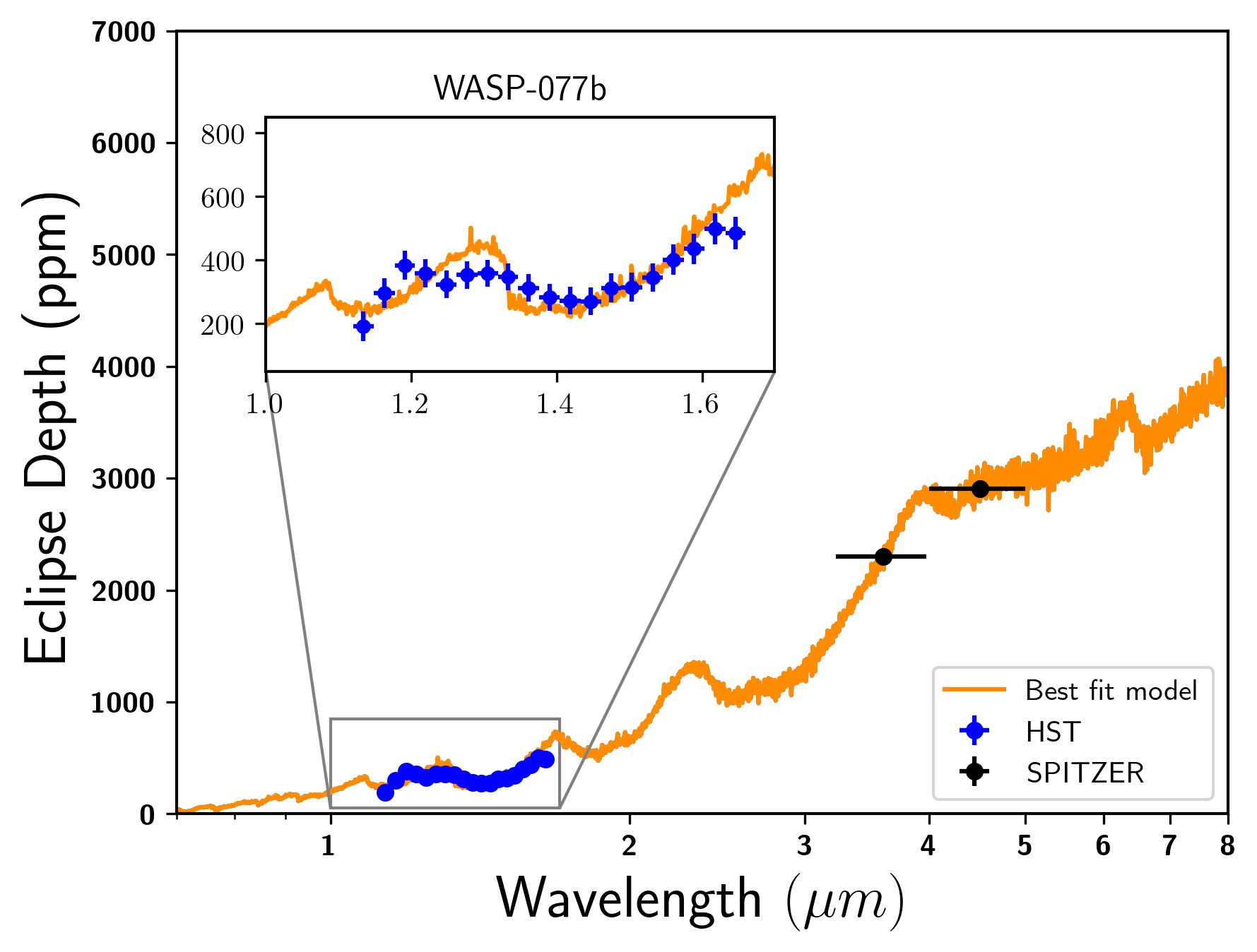}
    \caption{Best-fit emission spectrum model fitted to HST and \textit{Spitzer} observations of WASP-77b.}
    \label{fig:bestfit_wasp-77}
\end{figure}

WASP-77b was observed by TESS in sector 4 and 31 in 2018 and 2020, respectively. The full phase curve fit for WASP-77Ab that we obtain is shown in Figure \ref{fig:pcwasp-77}. For WASP-77Ab we constrained $R_p/R_s$ to be $0.1296\pm0.0011$ 
higher than $0.1193_{-0.0014}^{+0.0008}$ constrained by \citet{Wong_2020}. However, \citet{Wong_2020} only used data from sector 4. Previously, \citet{Reggiani2022} also used TESS data from sector 4 and 31, but to measure the rotational period of WASP-77A. They constrained R$_p$ to be $1.22_{-0.04}^{0.01}$ but by adopting transit depth from \citet{Maxted_2013} and did not report the transit depth and/or the $R_p/R_s$ they obtained. The TESS bandpass occultation depth of WASP-77Ab was constrained to $94_{-53}^{+62}$ ppm with 1.63 $\sigma$ detection significance much higher than $53_{-22}^{+32}$ ppm reported by \citet{Wong_2020}. The best fit self consistent model that we obtain by fitting to HST \citep{Mansfield_2022} and \textit{Spitzer} \citep{Garhart_2020} observations as shown in Figure \ref{fig:bestfit_wasp-77} has a RCF value of 0.8, metallicity of 0.063x solar (sub-solar), C/O ratio of 0.01 (sub-solar), with the reduced $\chi^2$ value of 1.37. The best-fit model sub-solar metallicity that we obtain is consistent with \citet{Line2021} using high-resolution spectroscopy and \citet{August_2023} using JWST observations. We also constrain C/O ratio to be highly sub-solar and $P$-$T$ profile to be without any thermal inversion as shown in Figure \ref{fig:pt}. This best fit model corresponds to a thermal occultation depth of $84^{+13}_{-35}$ ppm which is substantially higher than the Planck thermal occultation of 11 ppm. Using this model thermal occultation contribution, we constrain the geometric albedo of WASP-77b to be $0.017^{+0.126}_{-0.147}$. We do not compare the A$_g$ values that we obtain with \citet{Wong_2020}, as we measure a substantially higher occultation depth and we also use a self-consistent atmosphere model grid to constrain A$_g$, in contrast to a simple Planck function used by \citet{Wong_2020}.

\subsection{WASP-43b}

\begin{figure}
    \centering
    \includegraphics[scale=0.5]{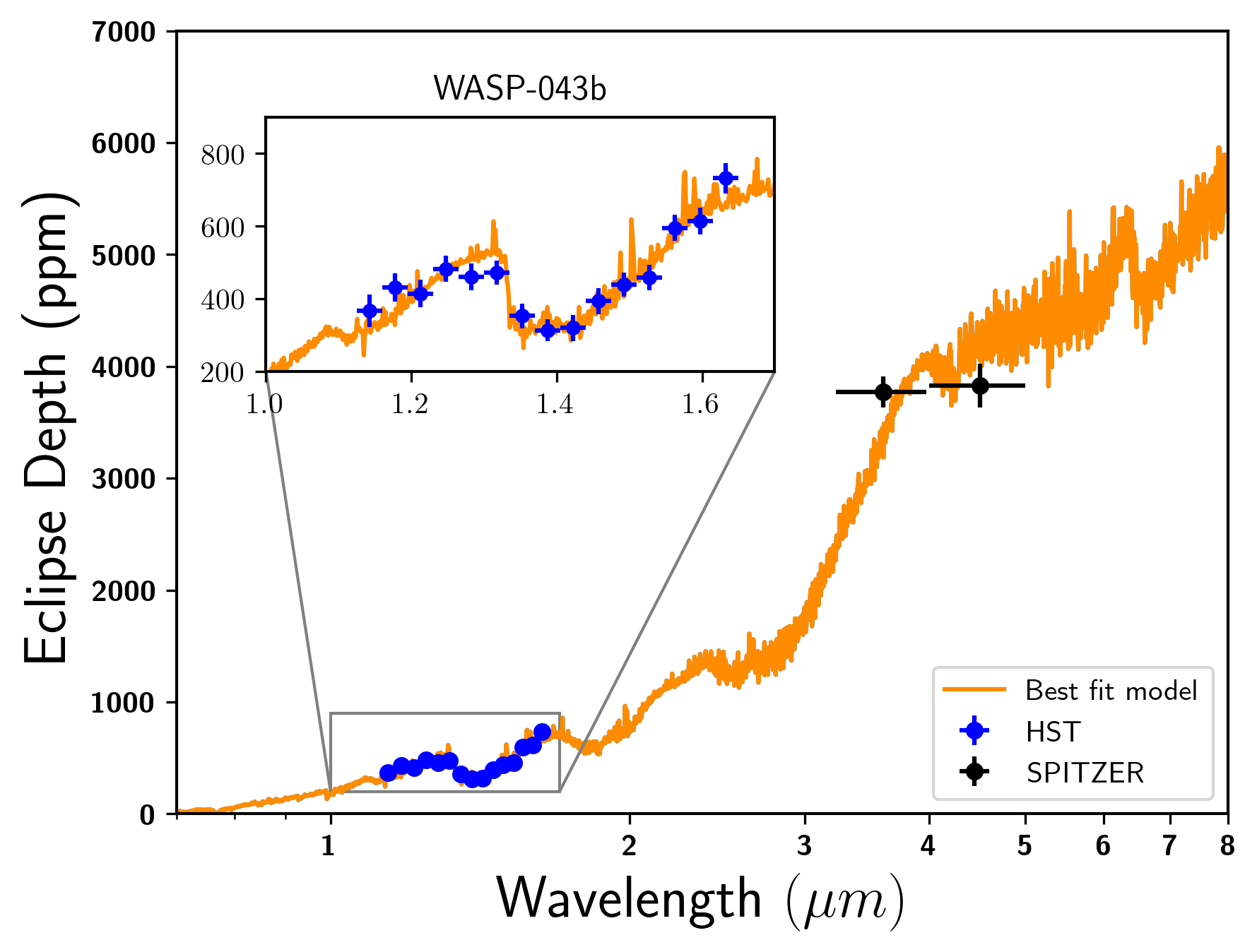}
    \caption{Best-fit emission spectrum model fitted to HST and \textit{Spitzer} observations of WASP-43b.}
    \label{fig:bestfit_wasp-43}
\end{figure}

WASP-43b was observed by TESS in sectors 9 and 35 in 2019 and 2021, respectively. We constrained its   $R_p/R_s$ to be $0.1580\pm0.0016$ with a transit depth of  $26.55\pm0.091$  ppt, consistent with $R_p/R_s$ of $0.15865\pm0.00044$ and transit depth of $26.597^{0.078}_{0.070}$ obtained by \citet{Bla_ek_2022}. The $(R_p/R_s)^2$ is slightly lower than the constrained transit depth due to the reasons discussed in section \ref{sec:benchmark}. We also constrain occultation depth to $182_{-56}^{+64}$ ppm with 3 $\sigma$ detection significance, which is higher than  $123_{-48}^{+59}$ ppm reported by \citet{Bla_ek_2022} and $70_{-40}^{+50}$ ppm reported by \citet{Scandariato2022}, but consistent with $170\pm70$ ppm reported by \citet{wong2020}. While \citet{wong2020} used data from only sector 9, \citet{Bla_ek_2022} and \citet{Scandariato2022} used data from sector 9 and 35 as we have done. The best-fit self-consistent model that we obtain for WASP-43b using its HST \citep{Kreidberg_2014} and \textit{Spitzer} \citep{Stevenson_2017} observations as shown in Figure \ref{fig:bestfit_wasp-43} has a RCF value of 0.8, metallicity of 0.01x solar, C/O ratio of 0.1, with the reduced $\chi^2$ value of 1.61, indicating good fit to the observations. While we constrain the metallicity to be highly sub-solar, \citet{Scandariato2022} constrains the metallicity to be solar ([M/H] = 0.1 $\pm$ 0.2). We also find the best fit $P$-$T$ profile to be without thermal inversion, as found by previous studies \citep{Kreidberg_2014}. With our best fit model we obtain the thermal contribution to the total occultation as $81^{+14}_{-23}$ ppm. With the reflected light component of 101 ppm we obtain geometric albedo of $0.109^{+0.086}_{-0.088}$ for WASP-43b, consistent with $0.116^{0.059}_{0.048}$ from \citet{Bla_ek_2022} but higher than the upper limit of 0.087 placed by \citet{Scandariato2022} and 0.06 placed by \citet{Fraine_2021}. The significant geometric albedo of WASP-43b that we obtain is contrasting to the results of previous studies, that constrain the atmosphere of WASP-43b to be cloudless \citep{Fraine_2021, Scandariato2022}.

\subsection{WASP-17b}

\begin{figure}
    \centering
    \includegraphics[scale=0.30]{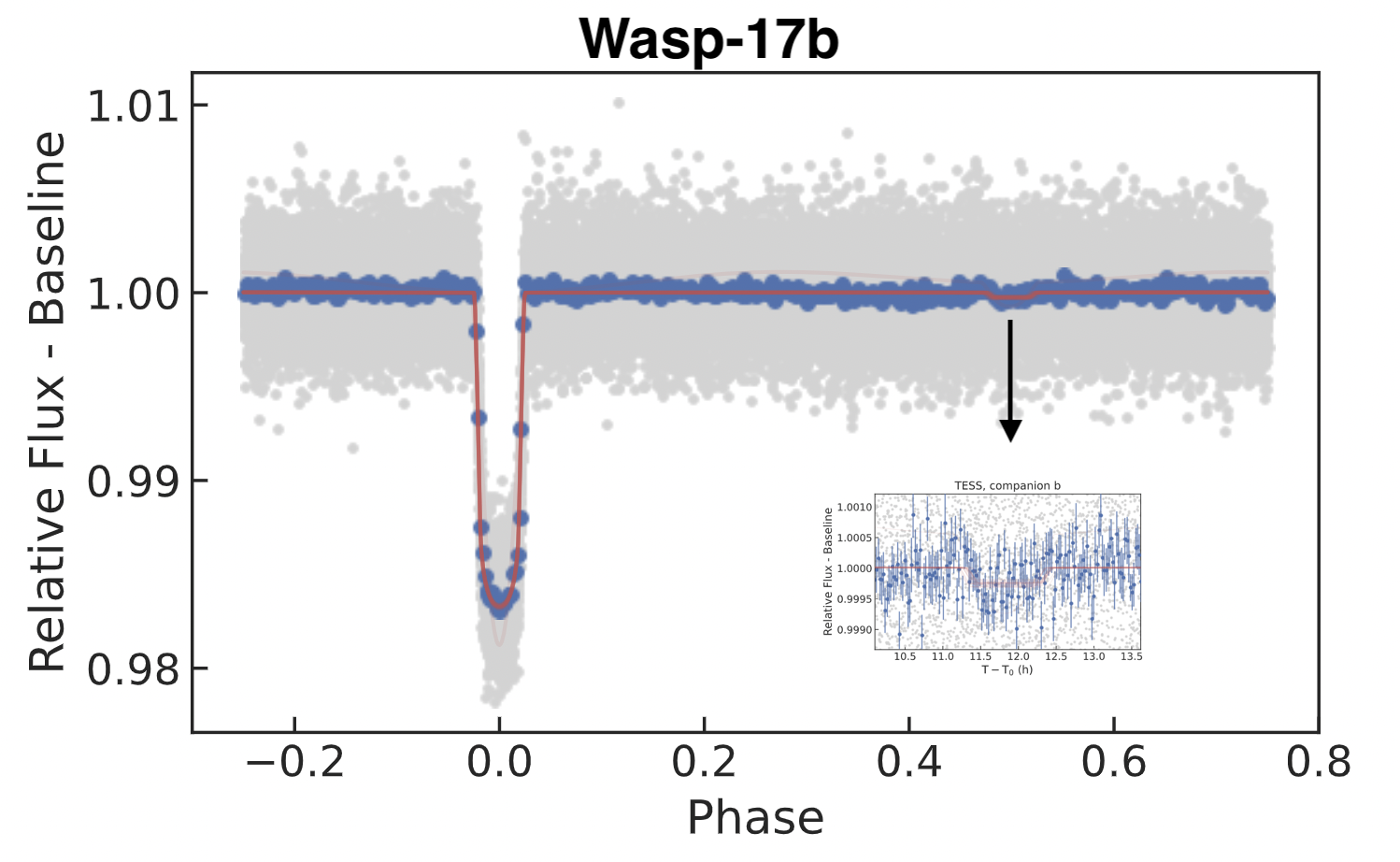}
    \caption{TESS full phase curve fit for WASP-17b.}
    \label{fig:pcwasp-18}
\end{figure}

\begin{figure}
    \centering
    \includegraphics[scale=0.5]{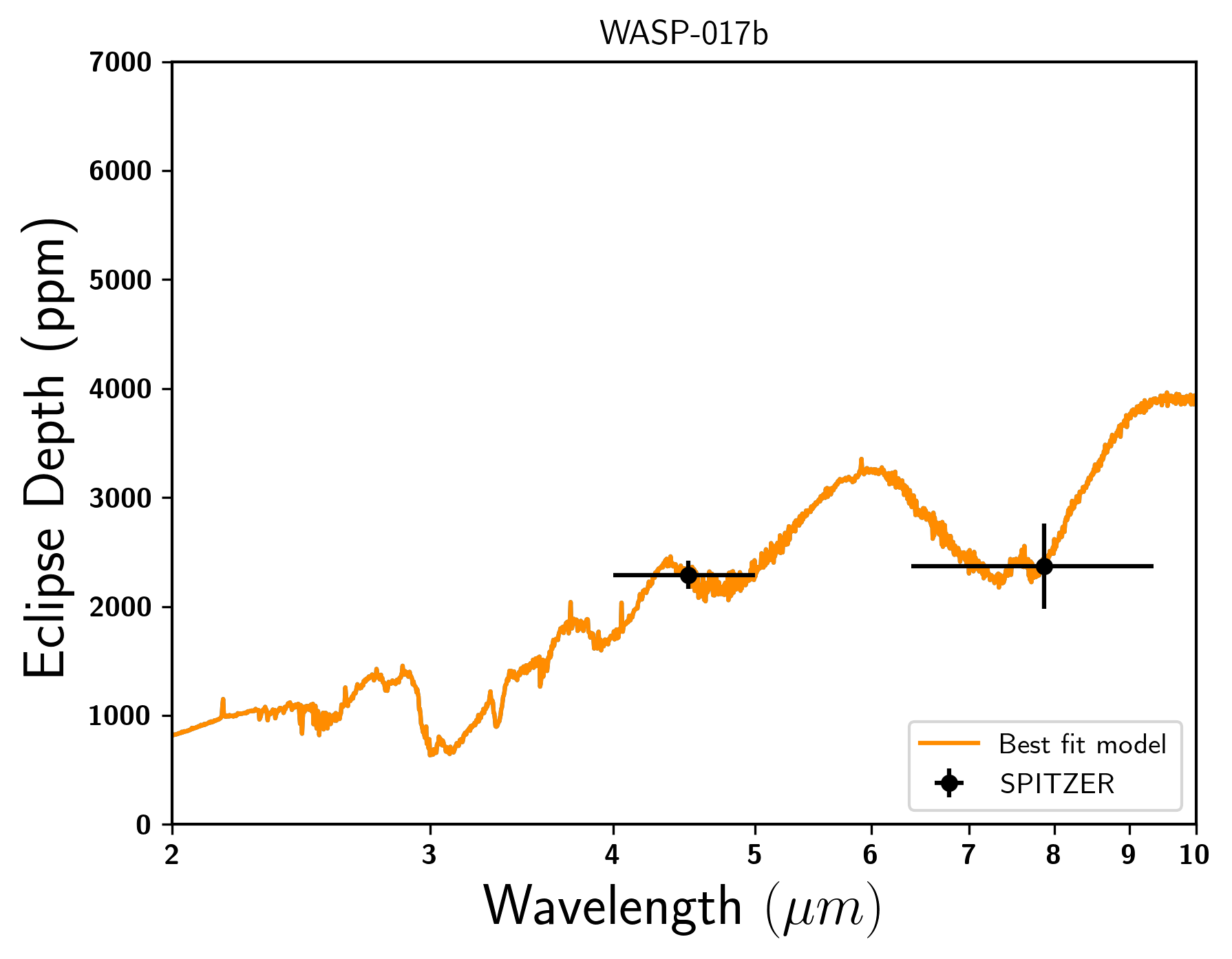}
    \caption{Best-fit emission spectrum model fitted to HST and \textit{Spitzer} observations of WASP-17b.}
    \label{fig:bestfit_wasp-17}
\end{figure}

WASP-17b had two sectors (12 and 38) of data available in the public domain which was used in this study. We constrained its planet to stellar radius ratio to be  $0.12382_{-0.00075}^{+0.00075}$, slightly higher than the square root of the white light curve transit depth, $0.12257_{-0.00011}^{+0.00011}$ constrained by \citet{Alderson2022} using HST STIS G750L grating, which almost covers similar band-pass as the TESS. We constrain the TESS measured occultation depth for WASP-17b to be $151_{-66}^{+83}$ ppm with 2 $\sigma$ detection significance . As far as authors are aware, this is the first estimate of the occulation depth of WASP-17b using TESS. The best-fit self-consistent model that we obtain for WASP-17b using its \textit{Spitzer} \citep{Anderson_2011} observations as shown in Figure \ref{fig:bestfit_wasp-17} has a RCF value of 0.8, metallicity of 2.5x solar, C/O ratio of 2.0, with the reduced $\chi^2$ value of 1.1, indicating good fit to the observations. Using this best fit model spectra we obtain a thermal contribution of just $34^{+67}_{-20}$  ppm in the TESS bandpass, leading to a substantial reflected component and therefore quite high albedo value of $0.401^{+0.526}_{-0.307}$. Even the lower limit on the albedo is greater than 0.094. This result hints at a comparatively higher reflective atmosphere than the other hot Jupiter exoplanets in our target list. The upcoming JWST observations of WASP-17b with NIRISS SOSS would be able to constrain this robustly. As for other planets, we obtain higher A$_g$ when we use Planck function model spectra.

\subsection{WASP-19b}

\begin{figure}
    \centering
    \includegraphics[scale=0.5]{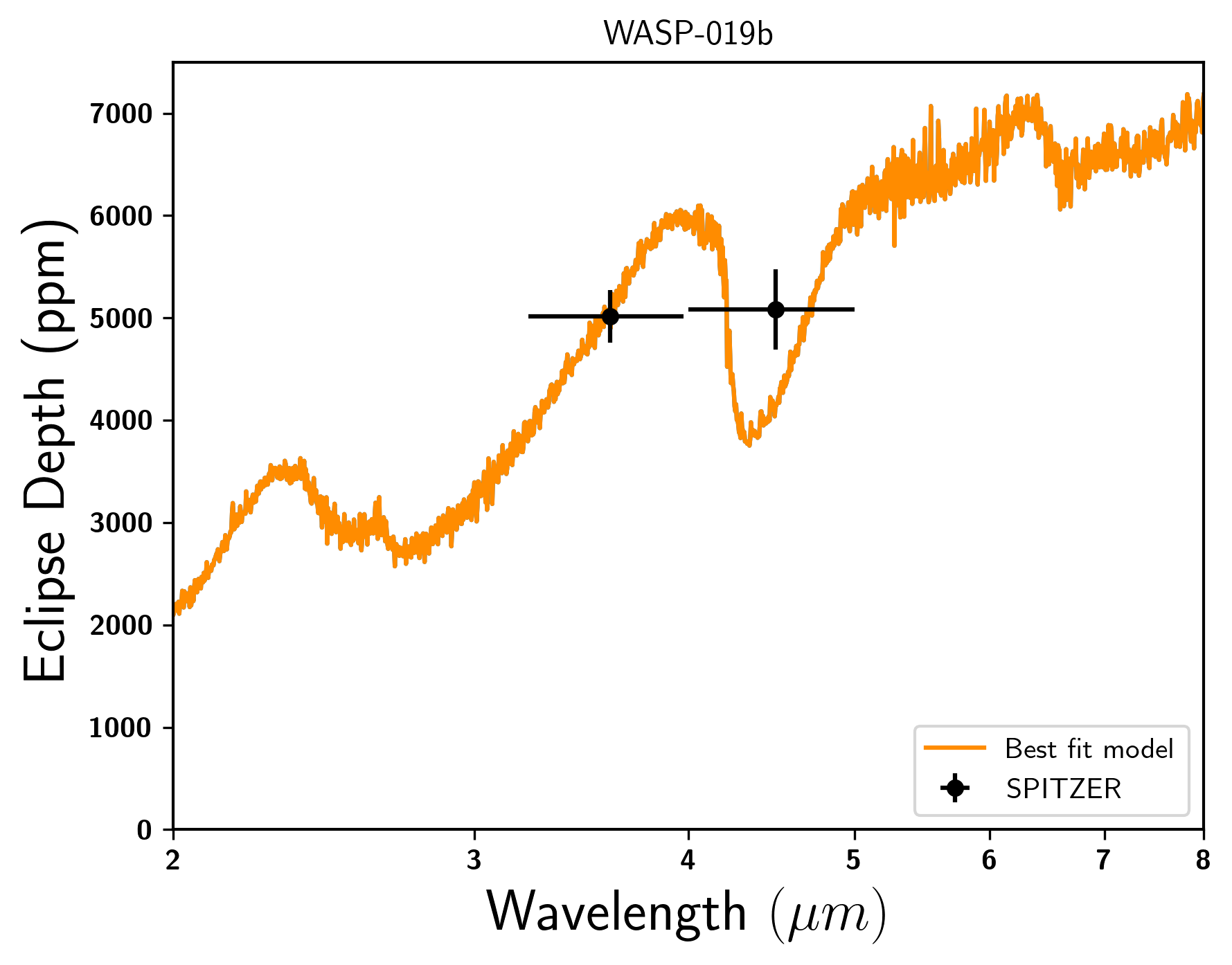}
    \caption{Best-fit emission spectrum model fitted to \textit{Spitzer} observations of WASP-19b.}
    \label{fig:bestfit_wasp-19}
\end{figure}

WASP-19b is a gas giant located about 260 pc away from earth. TESS observed it in sectors 9 and 36 in the year 2019 and 2021 with a total of 56 transits. The analysis of WASP-19b transit data was complicated and different because of WASP-19A being a very active star. \cite{Hebb_2009} reported a sinusoidal flux variation in the light curve of WASP-19 which constrained the rotational period of the star P$_{\rm rot}$ = 10.5 $\pm$ 0.2 days. Many studies such as \cite{Wong_2020} and \cite{Huitson_2013} incorporate effects of stellar variability using different methods. In this study we studied the two sector data separately. We studied the data without the inclusion of stellar variability and constrained the occultation depths from two different sectors to be 440 $\pm$ 100 ppm(sector 9) and 820$^{+130}_{-140}$ ppm(sector 36). Therefore to account for the stellar variability, the baseline (assumed to be a straight line) is fitted using red noise and Gaussian processes. With the best fit transit for sector 9 we obtain $R_p/R_s$ = 0.1468 $\pm$ 0.0015 which is slightly different from 0.15243$^{+0.00085}_{-0.00086}$ reported by \cite{Wong_2020}. Contrary to transit depth our occultation depth 484 $\pm$ 100 ppm is very close to 477$^{+120}_{-104}$ ppm published in \cite{Wong_2020}, when stellar variability is included. For Sector 36 we obtain $R_p/R_s$ = 0.1448$^{+0.0015}_{-0.0018}$ and an occultation depth of 570 $^{+120}_{-130}$. The combined data constrains $R_p/R_s$ to $0.1461\pm0.0013$  and  occultation depth to $549\pm89$ ppm with 6.1 $\sigma$ detection significance. Both these values for different sectors are within the error bars, after the inclusion of stellar variability. The best-fit model that we obtain by fitting to the \textit{Spitzer} \citep{Garhart_2020} observations of WASP-19b has RCF of 1.0, metallicity of 100x solar and sub-solar C/O ratio (0.01), with a reduced $\chi^2$ value of 0.025. Using this best-fit model we constrain the thermal contribution to the total occultation to be $601^{+0}_{-255}$, leading to very low geometric albedo (<0.022). The availability of only two data points for WASP-19b leads to poor constraints as well as very low reduced $\chi^2$ value. WASP-19A is a very active star and therefore exhibits large variations in its flux. Therefore, the recent \textit{Spitzer} observed data from \citet{Garhart_2020}, to which we fit our model grid, could have overestimated the errors for WASP-19b eclipse depth, leading to very low reduced $\chi^2$, unlike other planets in this work. Moreover, \citet{Garhart_2020} reported two different eclipse depths from the same observations of WASP-19b but with different bin sizes in their pixel level de-correlation method. The difference between the eclipse depth was quite substantial especially at 4.5 $\mu$m (5081 $\pm$ 392 ppm vs. 5848 $\pm$ 544 ppm ), thus highlighting the issues with the \textit{Spitzer} WASP-19b dataset, due to high stellar activity. We fitted our model grid to this other set of eclipse depth as well, but still obtained very low reduced $\chi^2$ values.

\begin{figure}
    \centering
    \includegraphics[scale=0.2]{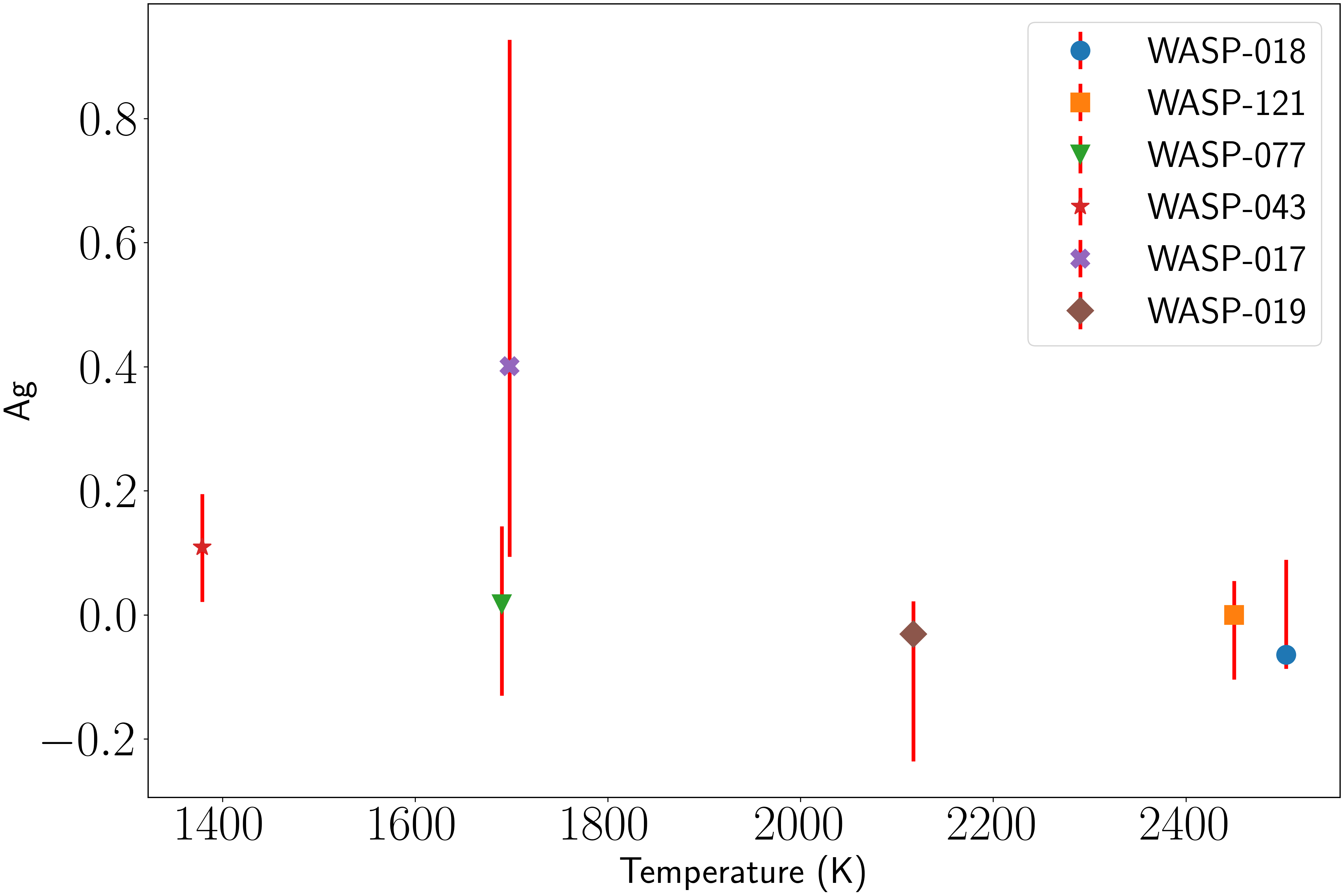} \\
    \includegraphics[scale=0.2]{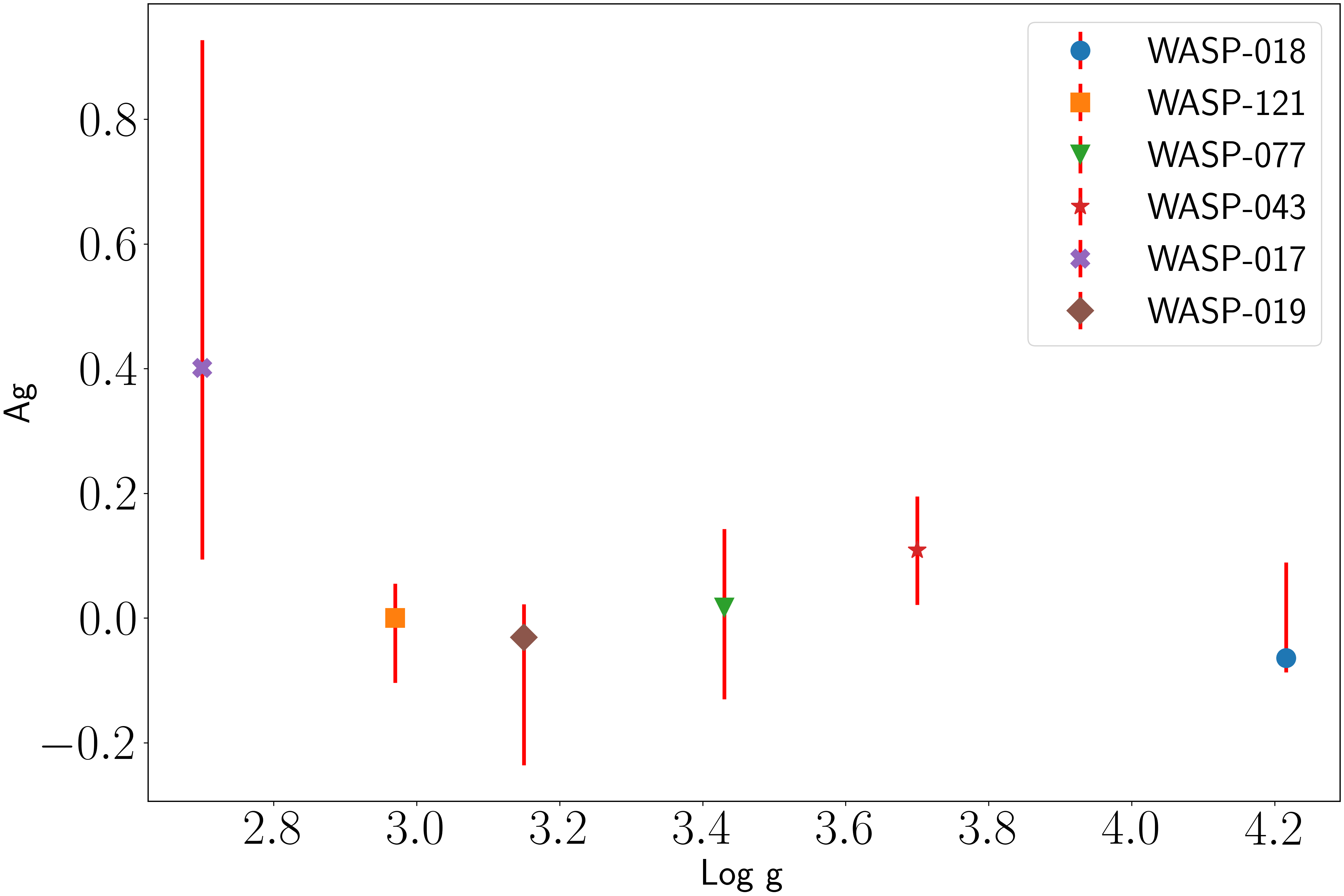}
    \caption{Figure showing geometric albedo (A$_g$) of target planets that we obtain in this work as a function of planetary equilibrium temperature (top figure) and Planetary gravity, i.e Log(g) (bottom figure)}
    \label{fig:Ag_temp_logg}
\end{figure}

\begin{table*}
\begin{tabular}{|l|l|l|l|l|l|l|} \hline 
\hline
Planet   & Total occultation   & Thermal Occultation & Thermal occultation & Ag & Ag      \\

&  & (Planck) & (Model) & (Planck) & (Model)  & \\ \hline

WASP-018 & $337^{+12}_{-12}$ (28 $\sigma$)                                & 107                        & $387^{+107}_{-6}$                       & 0.297            & <0.089                \\ \hline
WASP-121 & $405^{+22}_{-22}$   (18.4 $\sigma$)                                 & 121                        & $405^{+36}_{-88}$                       & 0.268            & $0.0^{+0.055}_{-0.104}$                   \\ \hline
WASP-077 & $94^{+62}_{-53}$   (1.63 $\sigma$)                                & 11                         & $84^{+13}_{-35}$                        & 0.138            & $0.017^{+0.126}_{-0.147}$                    \\ \hline
WASP-043 & $182^{+64}_{-56}$  (3 $\sigma$)                                & 5                          & $81^{+14}_{-23}$                       & 0.191            & $0.109^{+0.086}_{-0.088}$                   \\ \hline
WASP-017 & $151^{+83}_{-66}$       (2 $\sigma$)                          & 6                          & $34^{+67}_{-20}$                        & 0.495            & $0.401^{+0.526}_{-0.307}$   \\ \hline
WASP-019 & $549^{+89}_{-89}$  (6.1 $\sigma$)                                & 107                        & $601^{+0}_{-255}$                       & 0.264            & <0.022                  \\ \hline
\end{tabular}
\caption{Table showing TESS measured occultation depth (Column 2) and their detection significance, thermal contribution to occultation depth using Planck function (Column 3) and that due to self-consistent model grid with corresponding 1$\sigma$ condidence interval (Column 4), TESS bandpass geometric albedo derived using Planck function (Column 5) and that due to self-consistent model grid with 1$\sigma$ confidence interval (Column 6).}
\label{tab:albedo}
\end{table*}

\begin{table*}
\begin{tabular}{|l|l|l|l|l|l} \hline 
Planet & RCF  & M/H(x solar) & C/O & T$_{\rm int}$(K) & $\chi^2_{\rm red}$\\ \hline 
WASP-018&  1.2 & 0.63 & 0.55 &  400 & 1.28\\  \hline   
WASP-121 & 1.0 & 0.63 & 0.8 & 400 & 1.56\\ \hline   
WASP-077 & 0.8 & 0.063 & 0.01  & 100 & 1.37\\ \hline 
WASP-043 & 0.8 & 0.01 & 0.1 & 200 & 1.61\\ \hline   
WASP-017 & 0.8 & 2.5 & 2.0 & 100 & 1.1 \\ \hline   
WASP-019 & 1.0 & 100 & 0.01 & 100 & 0.025 \\ \hline
 
\end{tabular}
\caption{Table showing best-fit model parameters, re-circulation factor (RCF), metallicity (M/H), C/O ratio, internal temperature (T$_{\rm int}$) from the self-consistent atmosphere model grid for each planet using their HST and/or \textit{Spitzer} observations. The best-fit model reduced $\chi^2_{\rm red}$ value is also shown.}
\label{tab:model_fit_param}
\end{table*}

\begin{table*}
\begin{tabular}{|l|l|l|l|l|l|l|}
\hline
Fitted parameters                 & WASP-018                               & WASP-121                           & WASP-043                     & WASP-017                           & WASP-077                           & WASP-019                      \\ \hline
$R_b / R_\star$                   & $0.09768\pm0.00019$                    & $0.12128_{-0.00030}^{+0.00032}$    & $0.1580\pm0.0016$            & $0.12382_{-0.00075}^{+0.00075}$                & $0.1296\pm0.0011$                  & $0.1461\pm0.0013$             \\ \hline
$\cos{i_b}$                       & $0.1048\pm0.0049$                      & $0.045_{-0.015}^{+0.012}$          & $0.1406_{-0.0035}^{+0.0033}$ & $0.0674_{-0.0054}^{+0.0048}$       & $0.050_{-0.020}^{+0.015}$          & $0.1838_{-0.0059}^{+0.0052}$  \\ \hline
$q_{1; \mathrm{TESS}}$            & $0.219_{-0.019}^{+0.021}$              & $0.207_{-0.030}^{+0.033}$          & $0.46_{-0.14}^{+0.19}$       & $0.154_{-0.060}^{+0.090}$          & $0.214_{-0.079}^{+0.10}$           & $0.286_{-0.095}^{+0.13}$      \\ \hline
$q_{2; \mathrm{TESS}}$            & $0.310_{-0.036}^{+0.039}$              & $0.226_{-0.041}^{+0.047}$          & $0.31_{-0.18}^{+0.25}$       & $0.41_{-0.21}^{+0.30}$             & $0.43_{-0.15}^{+0.22}$             & $0.31_{-0.20}^{+0.33}$        \\ \hline
$J_\mathrm{b; TESS}$              & $0.0040\pm0.0012$                      & $0.0067_{-0.0013}^{+0.0014}$       & $0.0060\pm0.0024$            & $0.0172\pm0.0036$                  & $0.0053_{-0.0033}^{+0.0038}$       & $0.0044_{-0.0030}^{+0.0036}$  \\ \hline
$A_\mathrm{b; beaming; TESS}$     & $0.0312\pm0.0035$                      & $0.0215\pm0.0069$                  & $0.021_{-0.012}^{+0.014}$    & $0.0023_{-0.0017}^{+0.0038}$       & $0.057\pm0.021$                    & $0.145_{-0.091}^{+0.12}$      \\ \hline
$A_\mathrm{b; atmospheric; TESS}$ & $0.2981\pm0.0087$                      & $0.301\pm0.018$                    & $0.031_{-0.021}^{+0.028}$    & $0.0035_{-0.0026}^{+0.0054}$       & $0.0053_{-0.0040}^{+0.0088}$       & $0.45\pm0.11$                 \\ \hline
$A_\mathrm{b; ellipsoidal; TESS}$ & $0.3467\pm0.0088$                      & $0.0029_{-0.0022}^{+0.0050}$       & $0.106\pm0.035$              & $0.017_{-0.012}^{+0.021}$          & $0.105\pm0.050$                    & $0.037_{-0.027}^{+0.048}$     \\ \hline
$\Delta F_\mathrm{TESS}$          & $-0.0002413_{-0.0000072}^{+0.0000076}$ & $0.000016_{-0.000011}^{+0.000011}$ & $0.000150\pm0.000025$        & $0.000122_{-0.000015}^{+0.000014}$ & $0.000133_{-0.000029}^{+0.000031}$ & $0.0008_{-0.0070}^{+0.0064}$  \\ \hline
$\ln {\sigma_\mathrm{TESS}}$      & $-7.4153\pm0.0029$                     & $-6.7368\pm0.0029$                 & $-6.1121\pm0.0034$           & $-6.2372\pm0.0040$                 & $-6.2813\pm0.0058$                 & $-5.8343_{-0.0039}^{+0.0042}$ \\ \hline
gp: $\ln \sigma$                  & -                                      & -                                  & -                            & -                                  & -                                  & $-6.064_{-0.066}^{+0.069}$    \\ \hline
gp: $\ln \rho$                    & -                                      & -                                  & -                            & -                                  & -                                  & $-1.095\pm0.080$              \\ \hline
\end{tabular}
\caption{Table showing constraints on the fitted parameters obtained from TESS observations for each of our target planets using ALLESFITTER. Planet to star radius ratio ($R_b / R_\star$),Limb darkening parameters ($q_{1; \mathrm{TESS}}$ , $q_{2; \mathrm{TESS}}$), Cosine of inlcination angle ($\cos{i_b}$), Surface brightness ratio ($J_\mathrm{b; TESS}$), Doppler boosting ($A_\mathrm{b; beaming; TESS}$), Atmospheric modulation ($A_\mathrm{b; atmospheric; TESS}$), Ellipsoidal modulation ($A_\mathrm{b; ellipsoidal; TESS}$), log. error scaling ($\ln {\sigma_\mathrm{TESS}}$), Guassian parameters (gp: $\ln \sigma$ , gp: $\ln \rho$ ) }
\label{tab:allesfitter_fitted}
\end{table*}

\begin{table*}
\begin{tabular}{|l|l|l|l|l|l|l|}
\hline
Derived parameters                  & WASP-018                   & WASP-121                        & WASP-043                     & WASP-017                  & WASP-077                        & WASP-019                        \\ \hline
$R_\star/a_\mathrm{b}$              & $0.2853\pm0.0016$          & $0.2653_{-0.0022}^{+0.0026}$    & $0.2089_{-0.0026}^{+0.0028}$ & $0.1485\pm0.0022$         & $0.1890\pm0.0042$               & $0.2802\pm0.0031$               \\ \hline
$a_\mathrm{b}/R_\star$              & $3.505\pm0.019$            & $3.769_{-0.037}^{+0.031}$       & $4.786\pm0.063$              & $6.736\pm0.099$           & $5.29\pm0.12$                   & $3.569\pm0.039$                 \\ \hline
$R_\mathrm{b}/a_\mathrm{b}$         & $0.02787\pm0.00020$        & $0.03218_{-0.00033}^{+0.00038}$ & $0.03320\pm0.00058$          & $0.01838\pm0.00034$       & $0.02447_{-0.00067}^{+0.00072}$ & $0.04096_{-0.00067}^{+0.00059}$ \\ \hline
$R_\mathrm{b}$ ($\mathrm{R_{jup}}$) & $1.198\pm0.038$            & $1.752\pm0.071$                 & $1.042\pm0.017$              & $1.79\pm0.23$             & $1.285\pm0.090$                 & $1.436\pm0.045$                 \\ \hline
$a_\mathrm{b}$ (AU)                 & $0.02054\pm0.00067$        & $0.0260\pm0.0011$               & $0.01501\pm0.00029$          & $0.0467\pm0.0061$         & $0.0251\pm0.0018$               & $0.01677\pm0.00054$             \\ \hline
Inclination; $i_\mathrm{b}$ (deg)   & $83.98\pm0.29$             & $87.45_{-0.69}^{+0.89}$         & $81.92_{-0.19}^{+0.20}$      & $86.13_{-0.28}^{+0.31}$   & $87.13_{-0.85}^{+1.2}$          & $79.41_{-0.30}^{+0.35}$         \\ \hline
Impact parameter; $b$               & $0.371\pm0.016$            & $0.167_{-0.058}^{+0.044}$       & $0.673\pm0.011$              & $0.458_{-0.031}^{+0.027}$ & $0.268_{-0.11}^{+0.072}$        & $0.662_{-0.015}^{+0.012}$       \\ \hline
Equilibrium temperature (K)         & $2222\pm18$                & $2258_{-44}^{+48}$              & $1268_{-13}^{+15}$           & $1632\pm28$               & $1580\pm26$                    & $1922\pm25$                     \\ \hline
Transit depth (ppt)                 & $10.644_{-0.018}^{+0.016}$ & $16.409_{-0.035}^{+0.032}$      & $26.555\pm0.091$             & $16.715_{-0.11}^{+0.095}$ & $19.23_{-0.13}^{+0.12}$         & $22.35\pm0.12$                  \\ \hline
Occultation depth (ppt)             & $0.337\pm0.012$            & $0.405_{-0.022}^{+0.022}$       & $0.182_{-0.056}^{+0.064}$    & $151^{+83}_{-66}$ & $0.094_{-0.053}^{+0.062}$       & $0.549\pm0.089$                 \\ \hline
Nightside flux (ppt)                & $0.038\pm0.012$            & $0.098_{-0.018}^{+0.022}$       & $0.147\pm0.061$              & $0.264\pm0.057$           & $0.087_{-0.054}^{+0.062}$       & $0.089_{-0.065}^{+0.078}$       \\ \hline
\end{tabular}

\caption{Table showing constraints on the derived parameters obtained from TESS observations for each of our target planets using ALLESFITTER.}
\label{tab:allesfitter_derived}
\end{table*}

\section{Discussion}\label{sec:disc}
In the previous section we detailed the constraints on $R_p/R_s$, occultation depth, A$_g$ as well as the energy re-distribution, metallicity and C/O ratio that we obtain for each of our target planets. In this section, we discuss the trends that we see for the population of these exoplanets. 

We constrain A$_g$ using both Planck function and spectra generated using self-consistent atmosphere models. We find a general trend of Planck models overestimating the geometric albedos and underestimating the thermal occultation depth, when using the equilibrium temperature of the planet for computing the Planck models.  The model spectra generated using Planck function does not account for absorption/emission from the molecules present in the atmosphere and assumes a isothermal $P$-$T$ profile, which is a very crude assumption for obtaining realistic planetary emission spectra. The three hotter planets in our sample; WASP-18b, WASP-19b, and WASP-121b show the highest discrepancy between Planck and self-consistent model constraints on albedo. This is because equilibrium temperature computation assumes uniform re-distribution but these ultra-hot Jupiter planets tend to have almost no hear redistribution, as we see in the constraints on RCF shown in Table \ref{tab:model_fit_param} that we obtain for these planets.

In general, we find very low or negligible albedo values for the population of our targets, except for WASP-43b and WASP-17b. Equilibrium temperature and planet's gravity are two of the most important parameters that play a role in determining the albedo of a planet. Figure \ref{fig:Ag_temp_logg} shows the A$_g$ values of the target planets in this work, as a function of planet's equilibrium temperature and their gravity [log(g)], separately. When plotted as a function of T$_{\rm eq}$, we see two distinct set of regions, one below 2000\,K with non-zero A$_g$ values and one above 2000\,K with practically zero albedo in the TESS bandpass. When plotted as a function of log(g), WASP-17b stands out due to its comparatively low gravity and high albedo, however, WASP-43b even with its high gravity have significantly non-zero geometric albedo. This indicates a competition between T$_{\rm eq}$ and log(g) for determining A$_g$, potentially via cloud formation. The effects of temperature and gravity on cloud formation is well explained in \citet{Helling_2021, 2020A&A...634A..23W, Liu_2023, Estrela_2022}

\begin{figure}
    \centering
    \includegraphics[scale=0.22]{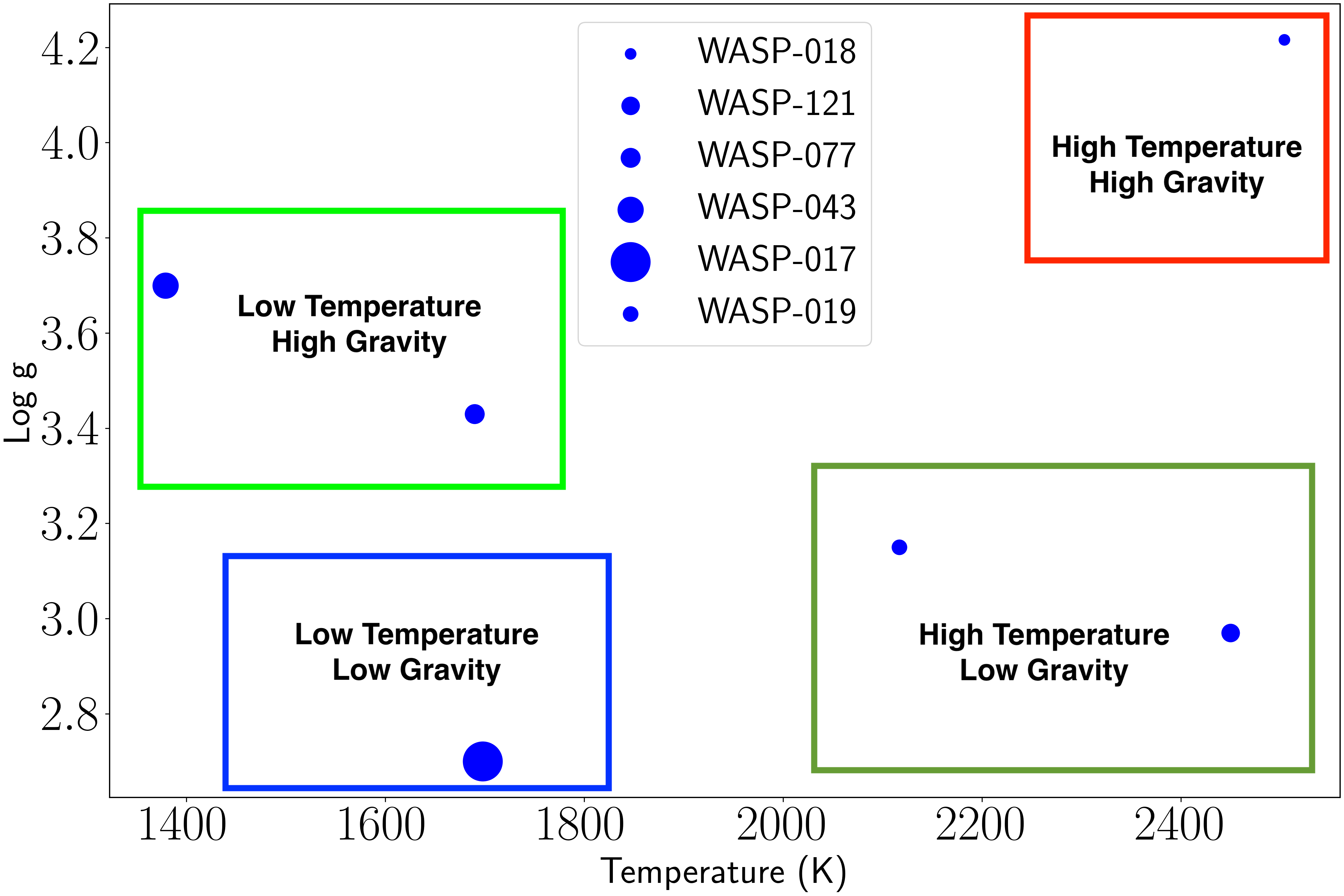}
    \caption{Figure showing the geometric albedo (A$_g$) values of the target planets (indicated by the sizes of the filled circles) as a function of equilibrium temperature and planet's Log(g) value.}
    \label{fig:tvsgvsAg}
\end{figure}

In Figure \ref{fig:tvsgvsAg} we show the A$_g$ values of the target planets in this work as a function of planet's equilibrium temperature and their gravity [log(g)], together. Low temperature and low gravity favor high albedo for a planet mostly because low temperature allows for easier condensation and formation of clouds whereas low gravity increases the cloud settling timescale leading to more cloud particles in the part of the atmosphere probed by our observations. An abnormally high value ($0.401^{+0.526}_{-0.307}$) of geometric albedo is observed for the case of WASP-17b which is consistent with the discovery of quartz clouds in WASP-17b \citep{grant2023jwsttst} in the JWST MIRI wavelength range (5 to 12 $\mu$m). The comparatively low equilibrium temperature of $\sim$1700\,K and low gravity of 5.49 m/s$^2$ for WASP-17b stands out to be the potential reason for this as shown in Figure \ref{fig:tvsgvsAg}. We also constrain a significant albedo ($0.109^{+0.086}_{-0.088}$) in the case of WASP-43 b. This value is 4 times lower than WASP-17b mostly because even with a lower temperature of just 1268\,K it has a significantly high gravity of about 50.93 m/s$^2$, almost 10 times that of WASP-17b. The low temperature condition will lead to easier formation of cloud but high gravity will lead to rapid settling and dissipation. On the other extreme we effectively constrain zero geometric albedo for WASP-18b with very high temperature and high gravity ($\sim$ 2700\,K and 164.43 m/s$^2$, respectively).

\section{Conclusion}\label{sec:conc}
In this work we measure the transit and occultation depth of six important JWST targets for atmospheric characterization, using their TESS Phase curve observations from multiple sectors. A dense grid of self-consistent model atmospheres (more than $\sim$ 8000 per planet) with radiative-convective equilibrium $P$-$T$ profiles consistent with equilbirum chemistry and their corresponding simulated emission spectra is generated. This grid is utilized to interpret the HST and/or the \textit{Spitzer} emission spectra observations of our target planets, constraining their energy re-distribution, metallicity and C/O ratio. The best-fit model spectra from the grid is utilized to compute the thermal contribution to the total occultation depth in the TESS bandpass and thereby constraints are obtained on the geometric albedo of the planet. Our important findings are as follows:

\begin{enumerate}
    \item We constrain a total occultation depth of $151_{-66}^{+83}$ ppm (2 $\sigma$) for WASP-17b in the TESS bandpass and a substantially high geometric albedo of  $0.401_{-0.307}^{+0.526}$ , in comparison to other hot Jupiter exoplanets with albedo constraints.  
 
     \item We constrain a total occultation depth of $182_{-56}^{+64}$ ppm  (3 $\sigma$) for WASP-43b in the TESS bandpass and a statistically signifcant geometric albedo of  $0.109_{-0.086}^{+0.088}$,  which is in tension with various studies constraining the dayside atmosphere of WASP-43b to be cloudless. JWST NIRISS SOSS observations of WASP-43b can shed more light on this inconsistency.
     
     \item We constrain a total occultation depth of $94^{+53}_{-62}$ ppm (1.63 $\sigma$) for WASP-77Ab, higher than previously published value, which used data from only one sector. The geometric albedo in the TESS bandpass was constrained to a negligible value of $0.017^{-0.126}_{+0.147}$. 
     
     \item  We find extremely low (practically zero) albedo values for WASP-18b (<0.089), WASP-19b (<0.022) and WASP-121b ($0.0^{+0.055}_{-0.104}$) using our self-consistent model grid, with their occultation depths consistent with previously published values.
     
     \item We constrain high re-circulation factor (low energy re-distribution) for all of our target planets, with WASP-018b having the lowest energy re-distribution.  
     
     \item We constrain sub-solar metallicity for WASP-18b, WASP-121b, WASP-77Ab and WASP-43b, with our self-consistent model grid best-fit to HST and \textit{Spitzer} observations.
     
     \item We constrain highly sub-solar C/O ratio for WASP-77Ab and WASP-43b, solar for WASP-18b and super-solar for WASP-121b,  with our self-consistent model grid best-fit to HST and \textit{Spitzer} observations.
     
     \item While looking at our target planets as a population, we find WASP-17b to be in the ideal place of low temperature and low gravity, conducive for cloud formation, possibly explaining its high geometric albedo.

     \item With our self consistent model grid best-fit to HST and \textit{Spitzer} observations, we constrain thermally inverted $P$-$T$ profiles for WASP-18b and WASP-121b, while $P$-$T$ profile for WASP-77b and WASP-43b do not show any thermal inversion.

\end{enumerate}

In this work, we have constrained the geometric albedo only in the  TESS bandpass. As discussed earlier, albedo is a wavelength dependant quantity, therefore wavelength dependant albedo constraints with JWST and/or WFIRST in the future can reveal a wealth of information including cloud and haze types, particle size distributions, etc., in these exoplanet atmospheres. In our current self-consistent models, we have not included clouds as we are only modelling the thermal contribution, however, we plan to do this in our future work. Additionally, the models used in this work are 1D, therefore further analysis with 3D models will be beneficial.

\section*{Acknowledgements}
We would like to thank the anonymous referee for their constructive comments and suggestions. We acknowledge support from the SERB SRG Grant SRG/2022/000727-G for this work. We would like to thank Prof. Subhasish Basak from NISER for the P452-Computational Physics course term project, that led to initiation of this work. 

\section*{Data Availability}

The TESS, HST and \textit{Spitzer} data used in this work are publicly available. The model data underlying this article will be shared on reasonable request to the corresponding author.



\bibliographystyle{mnras}








\bsp	
\label{lastpage}
\end{document}